\documentclass[apj]{emulateapj}
\usepackage{natbib}
\usepackage{graphicx}
\usepackage[toc,page]{appendix}
\usepackage{natbib}
\usepackage{color}
\usepackage{threeparttable}
\usepackage{hyperref}
\usepackage{breakurl}
\usepackage{amsmath}
\begin{document}

\slugcomment{Accepted for publication in the Astrophysical Journal}
\title{The Constant Average Relationship between Dust-obscured Star Formation and Stellar Mass from $z=0$ to $z=2.5$}
\email{kwhitaker@astro.umass.edu}
\author{Katherine E. Whitaker\altaffilmark{1,2,6}, Alexandra Pope\altaffilmark{2}, 
Ryan Cybulski\altaffilmark{3}, Caitlin M. Casey\altaffilmark{4}, Gerg\"{o} Popping\altaffilmark{5}, 
Min S. Yun\altaffilmark{2}} 
\altaffiltext{1}{Department of Physics, University of Connecticut, Storrs, CT 06269, USA}
\altaffiltext{2}{Department of Astronomy, University of Massachusetts, Amherst, MA 01003, USA}
\altaffiltext{3}{Department of Physics and Astronomy, Tufts University, Medford, MA 02155, USA}
\altaffiltext{4}{Department of Astronomy, The University of Texas at Austin, Austin, TX 78712, USA}
\altaffiltext{5}{European Southern Observatory, Karl-Schwarzschild-Strasse 2, D-85748, Garching, Germany}
\altaffiltext{6}{Hubble Fellow}
\shortauthors{Whitaker et al.}
\shorttitle{The Constant Average Relationship between Dust-obscured Star Formation and Stellar Mass}

\begin{abstract}

The total star formation budget of galaxies consists of the sum of the unobscured star
formation, as observed in the rest-frame ultraviolet (UV), together with the obscured component that is
absorbed and re-radiated by dust grains in the infrared. 
We explore how the fraction of obscured star formation depends on stellar
mass for mass-complete samples of galaxies at $0<z<2.5$.  We combine GALEX 
and WISE photometry for SDSS-selected galaxies with the 3D-HST treasury 
program and Spitzer/MIPS 24$\mu$m photometry in the well-studied 5 extragalactic CANDELS fields. 
We find a strong dependence of the fraction of obscured star formation ($f_{\mathrm{obscured}}$=SFR$_{\mathrm{IR}}$/SFR$_{\mathrm{UV+IR}}$) 
on stellar mass, with remarkably little evolution in this fraction 
with redshift out to $z$=2.5.
50\% of star formation is obscured for galaxies with log(M/M$_{\odot}$)=9.4; 
although unobscured star formation dominates the budget at lower masses, there exists a tail of 
low mass extremely obscured star-forming galaxies at $z>1$. 
For log(M/M$_{\odot}$)$>$10.5, $>$90\% of star formation is obscured at all redshifts.  
We also show that at fixed total SFR, $f_{\mathrm{obscured}}$ is lower at higher redshift.  
At fixed mass, high-redshift galaxies are observed to have more compact sizes 
and much higher star formation rates, gas fractions and hence surface densities 
(implying higher dust obscuration), yet
we observe no redshift evolution in $f_{\mathrm{obscured}}$ with stellar mass. This
poses a challenge to theoretical models to reproduce, where the observed compact 
sizes at high redshift seem in tension with lower dust obscuration.
\end{abstract}


\section{Introduction}
\label{sec:intro}

The current census of the peak epoch of star formation, $z\sim1-3$, 
reveals that the most massive galaxies are enshrouded in dust \citep[e.g.,][]{Magnelli09, Murphy11, Bourne17}.
These dust grains form a cocoon around nacent stars still embedded in their
native birth sites, preferentially absorbing the ultraviolet (UV) stellar radiation and thermally re-radiating
in the far-infrared (FIR) \citep{Seibert05, Cortese06, MunozMateos09}.  
It is this interstellar dust extinction that introduces the 
largest source of systematic error into global measurements of the star formation rates 
(SFR) of galaxies \citep{Kennicutt98}.  

There a several standard methods employed to account for dust attenuation
to attain a complete view of star formation.  
The unobscured star formation measured from the non-ionizing rest-frame UV continuum 
($\sim$130--250nm) of massive recently formed stars can be corrected using 
empirical calibrations based on the UV slope from the spectral energy distributions \citep[e.g.,][]{Meurer99}.  
However, these calibrations are uncertain due to potential intrinsic variations in the UV 
slope and dust attenuation curves \citep[e.g.,][]{Battisti16,Salmon16}. Alternatively, optical emission line diagnostics 
such as H$\alpha$ can be dust-corrected from the Balmer decrement \citep{Kennicutt09}. 
But such observations are difficult to obtain across cosmic time and observationally expensive.
One can also estimate total SFRs by directly measuring the 
obscured SFR$_{\mathrm{IR}}$ from one or more components of the mid- to FIR emission and co-adding this with the unobscured
UV SFR.  Although systematic uncertainties remain in calibrating the IR luminosity \citep{Calzetti13},
the main bottleneck has been in measuring accurate FIR emission for representative samples of 
star-forming galaxies across cosmic history.  

Due to the onset of source confusion, the deepest attainable FIR and sub-millimeter (sub-mm) 
surveys to date are only sensitive to the most extreme galaxies \citep[][and references therein]{Casey14a}. 
FIR and sub-mm selected galaxies are rare by number and likely do not represent the 
overall star-forming galaxy population. Given these current limitations, 
empirical calibrations of deep \emph{Spitzer}/MIPS and \emph{Herschel} stacks 
have instead been used to push the measurements of the average correlation between SFR and M$_{\star}$ down
to the equivalent M$_{\star}$ limits of the \emph{Hubble Space Telescope} (\emph{HST}) legacy 
surveys \citep{Viero13,Whitaker14b}.  Although the majority of star formation is assumed to be
unobscured in low-mass galaxies, the dust corrections are not insignificant.  
The goal of this paper is to quantify the level of obscured star formation 
as a function of M$_{\star}$ and total SFR for mass-complete samples of galaxies across
80\% of cosmic history (out to $z$=2.5).  We will address the following
questions: (1) how does the fraction of obscured star formation depend on M$_{\star}$ (Section~\ref{sec:mass})?, and 
(2) how does the fraction of obscured star formation depend on total SFR (Section~\ref{sec:sfr})?.  In Section~\ref{sec:discussion},
we discuss the redshift evolution of these relations in the context of our current understanding
of galaxy scaling relations and theoretical models.  The conclusions of this paper are summarized in Section~\ref{sec:conclusions}.

In this paper, we use a \citet{Chabrier} initial mass function (IMF) and assume a $\Lambda$CDM 
cosmology with $\Omega_{\mathrm{M}}=0.3$, $\Omega_{\Lambda}=0.7$, and $\mathrm{H_{0}}=70$ km s$^{-1}$ Mpc$^{-1}$. 

\section{Data and Sample Selection}
\label{sec:data}

\subsection{Stellar Masses, Redshifts and Rest-frame Colors}
We use the multi-wavelength datasets of five well-studied extragalactic 
fields (AEGIS, COSMOS, GOODS-N, GOODS-S, and UDS) through the Cosmic Assembly Near-IR Deep Extragalactic Legacy 
Survey \citep[CANDELS;][]{Grogin11} and the 3D-HST survey \citep{Momcheva15}.  
With M$_{\star}$, redshifts, and rest-frame colors from the 3D-HST 0.3--8$\mu$m photometric 
\citep{Skelton14} and spectroscopic \citep{Momcheva15} catalogs, we leverage the analysis of 
the mass-complete sample of 39,106 star-forming galaxies at $0.5<z<2.5$ presented in \citet{Whitaker14b}.  
The \citet{Skelton14} photometric catalogs include a large compilation of 
optical to near-infrared (NIR) photometric broadband and medium-bandwidth filters, 
ranging from 18 filters in UDS up to 44 in COSMOS.
Galaxies are identified using Source Extractor \citep{Bertin96} with deep
J$_{\mathrm{F125W}}$+H$_{\mathrm{F140W}}$+H$_{\mathrm{F160W}}$ detection images.
Redshifts and rest-frame colors are determined with the EAZY code \citep{Brammer08}.
Where possible, we combine the photometry with the spatially
resolved low-resolution HST/WFC3 G141 grism spectroscopy
to derive improved redshifts.
The ``best'' redshift was rank ordered to be spectroscopic (4\% total sample, 
13\% of galaxies with log(M$_{\star}$/M$_{\odot}$)$>$10), grism (12\% total, 38\% massive), or 
photometric (84\% total, 49\% massive), depending on the availability.
Star-forming galaxies are identified from rest-frame U-V and V-J colors following \citet{Whitaker12a}.
Luminous active galactic nuclei are identified and removed using the \emph{Spitzer}/IRAC
color selections of \citet{Donley12} due to its potential contamination of SFR$_{\mathrm{IR}}$.

Stellar masses M$_{\star}$ are derived by fixing the redshift to ``best'', as described above,
and fitting stellar population synthesis templates with the FAST code \citep{Kriek09b}. 
The FAST templates include a grid of \citet{BC03} models that assume a \citet{Chabrier} IMF 
with solar metallicity and a range of ages (7.6--10.1 Gyr), exponentially declining star
formation histories ($7<\tau<10$ in log years) and dust extinction
($0<A_V<4$). The dust content is parameterized by the attenuation in the V-band
following the \citet{Calzetti00} extinction law.
M$_{\star}$ is corrected for emission-line contamination of the 
broadband fluxes using values presented in Appendix A of \citet{Whitaker14b}.
The mass completeness limits employed herein correspond to the
90\% completeness limits derived by \citet{Tal14}, calculated
by comparing object detection in the CANDELS/deep with a
re-combined subset of the exposures that reach the depth of
the CANDELS/wide fields. Although the mass completeness
in the deeper GOODS-N and GOODS-S fields will extend to
lower stellar masses, we adopt the more conservative limits for
the shallower HST/WFC3 imaging.  The resulting sample is mass-complete down to
log(M/M$_{\odot}$) = 8.7 (9.3) at $z=1.0$ ($z=2.5$).

\subsection{Star Formation Rates}

SFR$_{\mathrm{IR}}$ originates from stacking analyses in
\citet{Whitaker14b}, using Spitzer/MIPS 24$\mu$m images 
from the Far-Infrared Deep Extragalactic Legacy survey \citep[AEGIS;][]{Dickinson07}, S-COSMOS survey 
\citep[COSMOS;][]{Sanders07}, GOODS Survey \citep[GOODS-N and GOODS-S;][]{Dickinson03}, and 
and the Spitzer UKIDSS Ultra Deep Survey\footnote{\url{http://irsa.ipac.caltech.edu/data/SPITZER/SpUDS/}}(UDS; PI:Dunlop).  
We use a high-resolution $J_{\mathrm{F125W}}$+$H_{\mathrm{F140W}}$+$H_{\mathrm{F160W}}$ 
detection image to model blended sources in the lower resolution 
MIPS/24$\mu$m image, ``cleaning'' all galaxies of the contaminating flux of 
neighboring sources \citep[Section 3,][]{Whitaker14b}. 
Although UV/optical emission may be spatially offset up to 
1$^{\prime\prime}$ from the IR emission \citep{Chen15,Koprowski16}, 
this is well within the 6$^{\prime\prime}$ MIPS/24$\mu$m beam.
The galaxy samples for four redshift intervals ($\bar{z}=0.75,1.25,1.75,2.25$, with $\Delta z=0.5$),
are sub-divided into bins of stellar mass with 0.2 dex width.  The number of galaxies
within a bin ranges from 7 upwards to 2498 galaxies, with an average value of $652\pm81$ galaxies. 
Unsurprisingly, the two bins with less than 20 galaxies represent the most massive 
galaxies (log(M$_{\star}$/M$_{\odot}$)=11.5) at the highest redshifts ($z>1.5$).
24$\mu$m flux densities are converted to total IR luminosity, L$_{\mathrm{IR}}$$\equiv$L(8--1000$\mu$m),
based on a single log average of the \citet[][hereafter DH02]{DH02} templates.
We explore this assumption in greater detail in Appendix A.
We convert L$_{\mathrm{IR}}$ to SFR$_{\mathrm{IR}}$ by multiplying by 
$1.09\times10^{-10}$  M$_{\odot}$ yr$^{-1}$ L$_{\odot}^{-1}$ \citep{Kennicutt98}, 
which assumes a \citet{Chabrier} IMF.
When considering only the sample above the SFR completeness limits, we confirm that 
the median stacked SFR$_{\mathrm{IR}}$ robustly probes the peak of the individual log-normal 
distributions within 0.05 dex \citep{Rodighiero11}.
Although the MIPS/24$\mu$m IR SFRs are generally robust for star-forming galaxies 
at these redshifts in aggregate \citep[e.g.,][]{Wuyts11a,Utomo14}, 
we compare the high-redshift results to \emph{Herschel}
and SCUBA-2 stacks from the literature in Section~\ref{sec:results}.

SFR$_{\mathrm{UV}}$ is derived from rest-frame UV luminosities based on 
\citet{Bell05}. 
The total integrated 1216--3000$\mathrm{\AA}$ UV luminosity is measured from the 2800$\mathrm{\AA}$ 
rest-frame luminosity multiplied by a factor of 1.5 to account for the UV 
spectral shape (L$_{\mathrm{UV}}$=1.5L$_{2800}$).
We adopt the L$_{2800}$ in lieu of 1600$\mathrm{\AA}$ to ensure that the UV
continuum is sampled by at least two photometric bands for all galaxies.  
L$_{2800}$ is determined from the best-fit template using
the same methodology as the rest-frame colors \citep{Brammer11}.  To derive
the SFR$_{\mathrm{UV}}$, \citet{Bell05} multiply L$_{\mathrm{UV}}$ by 
both $1.09\times10^{-10}$ M$_{\odot}$ yr$^{-1}$ L$_{\odot}^{-1}$ and
a factor of 2.2 that accounts for the unobscured starlight emitted shortward of 1216$\mathrm{\AA}$ and 
longward of 3000$\mathrm{\AA}$.  

\begin{figure}[b]
\leavevmode
\centering
\includegraphics[width=\linewidth]{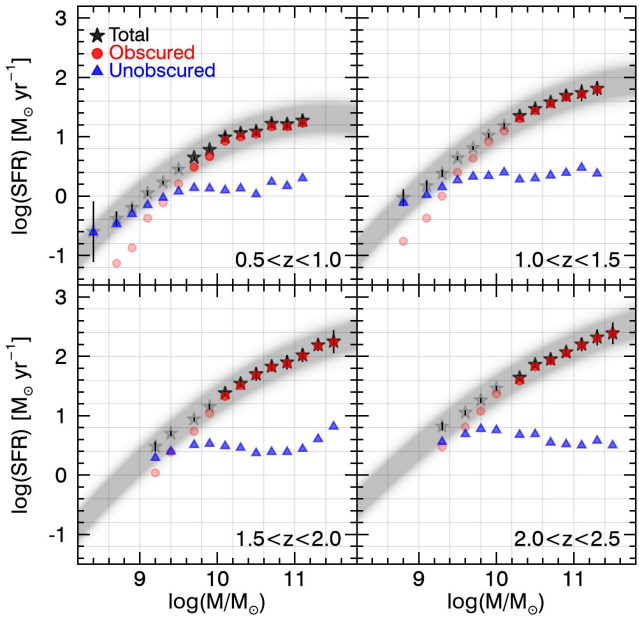}
\caption{The relation between total SFR and M$_{\star}$ (stars) is broken down into 
the respective unobscured (triangles) and obscured (circles) components. The grey
band with Gaussian shading corresponds to the typical 0.3 dex width of the observed relation 
\citep[e.g.][]{Whitaker12b, Speagle14}. The transparent points signify where SFR$_{\mathrm{IR}}$ (and
consequently SFR$_{\mathrm{total}}$) extend below the 3$\sigma$ limit for individual 
detections in the 24$\mu$m imaging; stacking to derive SFR$_{\mathrm{IR}}$ is necessary in this regime.
Above log(M$_{\star}$/M$_{\odot}$)=9.4 at 
all redshift epochs between $z$=0.5 and $z$=2.5, the majority of star formation is obscured.  Conversely, at lower
M$_{\star}$, star formation is unobscured on average.}
\label{fig:sfr_mass}
\end{figure}

Another popular alternative conversion to total SFR is presented in \citet{Murphy11}.
This conversion is not only different in the absolute calibration,
but also in the relative amount of star formation attributed to the obscured and unobscured phases.
Although we adopt the aforementioned SFR calibration to 
remain consistent with our previous work, we consider in Section~\ref{sec:mass} how this 
alternative calibration would systematically change the observed correlations.  

\begin{figure*}[t]
\leavevmode
\centering
\includegraphics[width=\linewidth]{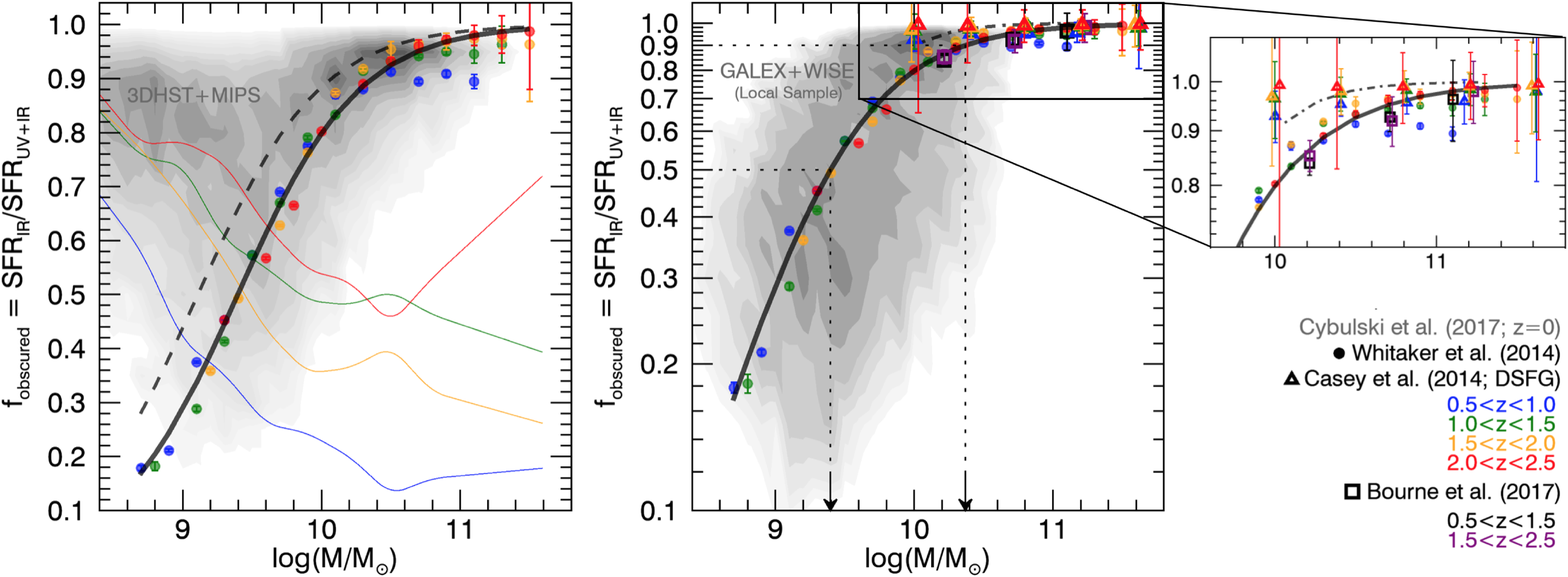}
\caption{The fraction of obscured star formation relative to total derived from median stacks 
is a strong function of M$_{\star}$ and does not evolve strongly out to $z$=2.5.
(Left) Linear y-axis version, including greyscale contours of the individual 
3D-HST measurements in addition to the median stacks (circles) from \citet{Whitaker14b}.  
The thin lines represent the completeness limits of the 24$\mu$m imaging. The dashed line is the average relation when 
using the \citet{Murphy11} SFR calibrations. 
(Right) Logarithmic y-axis version, comparing to a local
SDSS sample based on GALEX/WISE photometry at 0.02$<$$z$$<$0.05 in greyscale (Cybulski et al. in prep).  
We additionally compare to 450$\mu$m and 850$\mu$m SCUBA-2 
stacks of a mass-selected sample from \citet{Bourne17}, finding excellent agreement at the massive end (squares). 
The \emph{Herschel} imaging 
from \citet{Casey14} agrees at the massive end down to the limits of the FIR-selected sample 
of star-forming galaxies (triangles). The dot-dash line represents the implied ratio based
on UV light attenuation measurements by \citet{Pannella09}.}
\label{fig:frac_obscured}
\end{figure*}

\section{Results}
\label{sec:results}
\subsection{How does the fraction of obscured star formation depend on stellar mass?}
\label{sec:mass}

In Figure~\ref{fig:sfr_mass}, we present the median relation between SFR$_{\mathrm{UV+IR}}$ and M$_{\star}$ from 
$z$=0.5 out to $z$=2.5 from \citet{Whitaker14b}.  
As described in \citet{Whitaker14b}, the error bars for L$_{\mathrm{IR}}$ are
derived from a Monte Carlo bootstrap analysis of the 24$\mu$m stacks for each stellar mass bin,
and added in quadrature to the 1$\sigma$ scatter in the L$_{\mathrm{UV}}$ distributions.
We break down the total SFR into the 
respective unobscured (UV; blue triangles) and obscured (IR; red circles) components.  
The high transparency symbols signify SFR$_{\mathrm{IR}}$ below the 3$\sigma$ 24$\mu$m limit,
whereas more opaque symbols are above this limit.  While we adopt the stacked SFR$_{\mathrm{IR}}$ values 
for all bins, we note that the stacks are in agreement with the detections above the 3$\sigma$ 24$\mu$m limits.
At all redshifts considered, $>$50\% of the star formation is obscured for galaxies more massive
than log(M$_{\star}$/M$_{\odot}$)=9.4.  SFR$_{\mathrm{UV}}$ is relatively flat with increasing
M$_{\star}$ at each epoch, whereas star formation is increasingly obscured as galaxies gain
more stellar mass.  This figure demonstrates the importance of using the appropriate SFR
diagnostics depending on the M$_{\star}$ regime being considered; it is critical to obtain
IR observations to measure robust SFRs for massive galaxies, whereas rest-UV observations
will be more important at log(M$_{\star}$/M$_{\odot}$)$\lesssim$9.

\begin{table}[t]
\centering
\begin{threeparttable}
    \caption{Logistic Growth Function: $f_{\mathrm{obscured}}$ vs. log(M$_{\star}$/M$_{\odot}$)\label{tab:mass_fit}}
    \begin{tabular}{lcc}
      \hline \hline
      & {$a$} & {$b$} \\
      \hline
      \noalign{\smallskip}
      Standard Calibration  &  (1.96$\pm$0.14)$\times$10$^{9}$ &  -2.277$\pm$0.007 \\
      Murphy et al. (2011)  & (1.09$\pm$0.09)$\times$10$^{9}$  &		-2.284$\pm$0.009 \\
      \noalign{\smallskip}
      \hline
      \noalign{\smallskip}
    \end{tabular}
    \begin{tablenotes}
      \small
    \item \emph{Notes.} Logistic growth function coefficients parameterizing the evolution of 
    	$f_{\mathrm{obscured}}$-$\log($M$_{\star}$/M$_{\odot}$) using Equation~\ref{eq:mass_fit}.  
    		``Standard calibration''
    		corresponds to the adopted luminosity to SFR conversions described herein (solid line, Figure~\ref{fig:frac_obscured}); 
    		alternative \citet{Murphy11} conversion also parameterized (dashed line, Figure~\ref{fig:frac_obscured}).
    \end{tablenotes}
  \end{threeparttable}
\end{table}

We observe a remarkably tight correlation between the median obscured star formation fraction
($f_{\mathrm{obscured}}\equiv\widetilde{\mathrm{SFR}}_{\mathrm{IR}}/(\widetilde{\mathrm{SFR}}_{\mathrm{UV}}+\widetilde{\mathrm{SFR}}_{\mathrm{IR}})$) and M$_{\star}$, 
generally showing little redshift evolution between $z$=0.5 and $z$=2.5 (Figure~\ref{fig:frac_obscured}).
Here, we use the error analysis presented in Figure~\ref{fig:sfr_mass} in conjunction with the number of galaxies
within a particular bin to derive the error in the mean.
The best-fit relation is defined by a logistic growth function: 

\begin{equation}
f_{\mathrm{obscured}}=\frac{1}{1+a e^{b\log(\mathrm{M}_{\star}/\mathrm{M}_{\odot})}}.
\label{eq:mass_fit}
\end{equation}

The greyscale in the left panel of Figure~\ref{fig:frac_obscured} shows the contours of the individual 
3D-HST detections relative to the completeness
limits at each redshift epoch (thin solid lines).  Owing to the extremely deep optical
CANDELS/3D-HST photometry, the limiting factor for individual SFRs is the \emph{Spitzer}/MIPS 24$\mu$m depth.
We convert 3$\sigma$ 24$\mu$m limits
into the limiting SFR$_{\mathrm{IR}}$ for each redshift interval \citep[see Figure 2 in][]{Whitaker14b}.  
When combined with the SFR$_{\mathrm{UV}}$ in the 95th percentile for a given M$_{\star}$ and redshift bin, 
this yields an effective completeness limit. The shape of the completeness curves therefore depends on both
the redshift evolution in the distribution of SFR$_{\mathrm{UV}}$ and the limiting SFR$_{\mathrm{IR}}$.
The 95th percentile is chosen to avoid extreme outliers in SFR$_{\mathrm{UV}}$,
while probing the minimum obscuration fraction for a given SFR$_{\mathrm{IR}}$ limit.  

The scarcity of individual detections near the completeness limits for massive galaxies indicates 
that they are preferentially highly dust obscured; there is a dearth of massive galaxies that are relatively
unobscured. Galaxies are on average $>$95\% obscured at and above the knee of the mass function 
\citep[e.g., log(M$^{\star}$/M$_{\odot}$)=10.8--11,][]{Muzzin13}, 
and $>$70\% obscured 1 dex below the characteristic mass \citep[see also][]{Dunlop17}.
While the majority of star formation in the 
median low-mass galaxy will be observable in the rest-UV, there exists a population of highly obscured low-mass galaxies
at $z$$>$1 \citep[e.g.,][]{Pope17}.  The individual greyscale 3D-HST distribution relative to the median
stacked values suggests a large scatter in the amount of obscuration in galaxies with
log(M$_{\star}$/M$_{\odot}$)$<$9.5 at $0.5<z<2.5$.  

Next, we compare with results at $z\sim0$ (Figure~\ref{fig:frac_obscured}, right). The greyscale in the right panel represents 22,481
star-forming galaxies selected at $0.02<z<0.05$ from the Sloan Digital Sky Survey DR12 \citep[SDSS;][]{Alam15}. The
SFRs for this local comparison sample are measured from GALEX and WISE/22$\mu$m photometry (see
Cybulski et al. in prep for the details). The SDSS mass completeness sets in
at much higher masses (log(M/M$_{\odot}$)$\sim$10) than for 3D-HST. Close to these limits, the data also become limited
by the depth of the GALEX and WISE all-sky (but relatively shallow) photometry.  We require that the
FUV GALEX exposure time is $>100$ seconds, resulting in $>$80\% completeness across the redshift range. 
Quiescent galaxies are excluded from this sample on the basis of the specific SFR (SFR/M$_{\star}$) bimodality, 
consistent with the 3D-HST UVJ-selection, where we identify and remove galaxies with log(sSFR)$<$10$^{-11}$ yr$^{-1}$. 
To test how sensitive the resulting SDSS distribution is to this assumption, we conservatively 
raise and lower this limit in log(sSFR) by 0.5 dex.  Lowering the cut to log(sSFR)=-11.5 yr$^{-1}$ intersects 
the peak of the quiescent distribution, whereas raising it to log(sSFR)=-10.5 yr$^{-1}$ corresponds 
roughly to the 1$\sigma$ lower envelope of the star-forming galaxy distribution.  We find that 
the results are not sensitive to our definition of quiescence in the SDSS sample, with the median of 
the distribution changing only weakly ($\Delta\tilde{f}_{\mathrm{obscured}}<0.04$) for a correspondingly 
large change in the log(sSFR) limit.

The mode of the SDSS distribution in Figure~\ref{fig:frac_obscured} (right) tracks
the higher redshift data well, but with a broader distribution.  
In Figure~\ref{fig:frac_obscured}, there appears to be a dearth of highly obscured low-mass galaxies at $z\sim0$.  
We forgo interpretation of the individual distributions 
at the lowest M$_{\star}$ due to complications by the incompleteness limits of both data sets.
For obscuration fractions to the left of the dotted lines, SFR completeness effects 
will become important.

\begin{figure}[t]
\leavevmode
\centering
\includegraphics[width=0.98\linewidth]{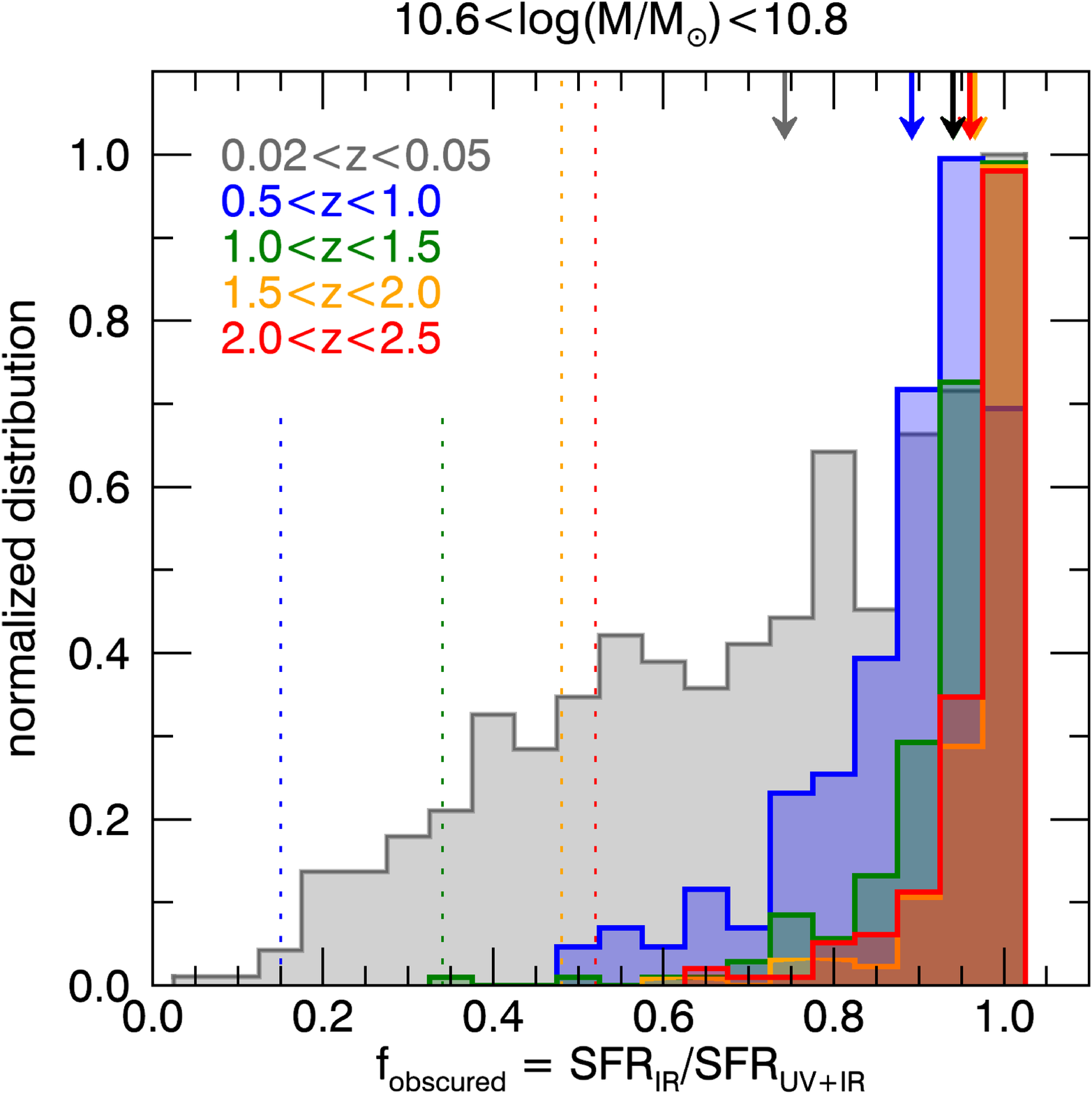}
\caption{Distributions of obscured fraction for 
individually detected galaxies with 10.6$<$log(M/M$_{\odot}$)$<$10.8 at 0.02$<$$z$$<$0.05 (SDSS; Cybulski et al. in prep)
and 0.5$<$$z$$<$2.5 \citep[3DHST;][]{Whitaker14b}.  The arrows mark the distribution medians and
the dotted lines (color-coded by redshift) indicate where samples are mass and SFR complete.  The black
arrow is the best-fit relation from Figure~\ref{fig:frac_obscured}.  While there is little to no evolution in
the mode, the width of the distribution increases towards lower redshifts. This bin is representative of the trends
for all M$_{\star}$ where the SDSS and 3D-HST samples are complete.}
\label{fig:frac_obscured_hist}
\end{figure}

We can illuminate the difference in 
the distributions from $z$=2.5 to $z$=0 for massive galaxies by selecting a bin 
of 10.6$<$log(M/M$_{\odot}$)$<$10.8, where both the SDSS and 3D-HST samples 
are complete down to low levels of obscured star formation (Figure~\ref{fig:frac_obscured_hist}).  
We find that the distribution at $z$=0 monotonically increases towards a maximum value at 100\% obscuration.  
The mode of the distributions (95-100\%) remains relatively unchanged out to $z$=2.5, whereas
the width narrows.  Although we only show one M$_{\star}$ bin here, we find similar trends down to log(M/M$_{\odot}$)=10.
When measuring the median of the 
distributions (arrows), as also done in the stacking analysis, this suggests a decrease 
(or flattening) in obscuration at low-redshift for the most massive galaxies.
This is also seen in the left panel of Figure~\ref{fig:frac_obscured}, where there exists a 
noteworthy deviation from the best-fit relation for the most massive galaxies in 
the $0.5<z<1.0$ ($1.0<z<1.5$) redshift bin, with $f_{\mathrm{obscured}}$ for galaxies with log(M$_{\star}$/M$_{\odot}$)=11 
depressed by 10\% (4\%) or 0.05 dex (0.02 dex); see also \ref{fig:frac_obscured_templates}).
As shown in \citet{Kauffmann03}, the vast majority of the most massive galaxies in SDSS 
at $z=0$ are quiescent.  
This makes it difficult to push our direct comparison of the distributions presented in 
Figure~\ref{fig:frac_obscured_hist} towards the highest stellar masses, owing to the sharp 
drop off in the local sample.

The observation that star formation in the most massive galaxies 
(log(M/M$_{\odot}$)$>$10.6) at intermediate redshifts is slightly less
dust-obscured can also be seen in Figure 5 of \citet{Whitaker14b}, who show 
IRX ($\equiv$L$_{\mathrm{IR}}$/L$_{\mathrm{UV}}$) for 
this same data analysis as a function of stellar mass.  This same redshift bin also has a lower IRX-$\beta$
relation \citep[Figure 6][]{Whitaker14b}; the results presented in the right panel 
of Figure~\ref{fig:templates} that adopt an 
evolving template conversion for L$_{\mathrm{IR}}$ suggest that we are only under-estimating L$_{\mathrm{IR}}$ when
using the DH02 log-average template for the most IR-luminous galaxies.  The choice of template therefore
will not significantly reduce discrepancies in the
IRX-$\beta$ relation or $f_{\mathrm{obscured}}$, as we are probing the median properties of the star-forming 
galaxy population where the ratio of L$_{\mathrm{IR}}$($z$)/L$_{\mathrm{IR,DH02}}$ is approximately unity.
We explore template-dependent effects on L$_{\mathrm{IR}}$ in greater detail below and in Appendix \ref{sec:appendixA}.
Both a relative increase in far-UV compared to near-UV attenuation and an increasing stellar population age
could result in the shallower IRX-$\beta$ and $f_{\mathrm{obscured}}$-log(M$_{\star}$) relations 
observed \citep[see e.g., Figure 11 in][]{Popping17b}.

\subsubsection{The Effects of Alternative SFR$_{\mathrm{IR}}$ Methodologies}
\label{sec:sfrirmethod}

There exists a range of alternative SFR$_{\mathrm{IR}}$ methodologies adopted within the literature.
We will explore how our results depend on the assumed calibrations and templates in the following paragraphs.
The dashed line in Figure~\ref{fig:frac_obscured} (left) corresponds to the best-fit relation when 
corrected to the SFR calibrations presented in Equations 3 and 4 of \citet{Murphy11}.  
Whereas the relative ratio of total star formation originating in IR relative to UV emission 
is assumed to be SFR$_{\mathrm{IR}}$/SFR$_{\mathrm{UV}}$=0.45 here, \citet{Murphy11} adopt
a higher value of SFR$_{\mathrm{IR}}$/SFR$_{\mathrm{UV}}$=0.88.  This results in an overall shift
of $f_{\mathrm{obscured}}$ towards higher fractions of obscured star formation.

Studies show that the MIPS/24$\mu$m photometry is a robust tracer of the average 
SFR$_{\mathrm{IR}}$ out to $z\sim2-3$ \citep[e.g.,][]{Wuyts11a, Tomczak16}. 
However, to rule out potential biases associated with the 24$\mu$m empirical SFR calibration, 
we compare our results to several independent studies.  \citet{Bourne17} present a stacking
analysis based on the ultra-deep 450$\mu$m and 850$\mu$m over 230 arcmin$^{2}$ from the
SCUBA-2 Cosmology Legacy Survey in the AEGIS, COSMOS and 
UDS fields, together with 100-250$\mu$m imaging from \emph{Herschel}. They adopt a similar deblending
approach for the longer wavelength data that relies on the deep photometric catalogs from 
CANDELS/3D-HST. We adopt a correction to the stellar masses presented in \citet{Bourne17} 
of -0.03 dex to convert from Kroupa to Chabrier IMF, following \citet{Zahid12}. 
We further correct their adopted luminosity 
to SFR conversion of \citet{Murphy11} to that used in the present analysis.
The utility of the deep MIPS/24$\mu$m photometry is evident when considering the 
M$_{\star}$ limits achievable with their data analysis.  Though the average trends
are in excellent agreement with our results, \citet{Bourne17} are only probing the most massive, obscured galaxies
just shy of the M$_{\star}$ regime where the trend sharply falls. They push
below the standard confusion limit, but the data is not able to probe the full dynamic range in M$_{\star}$.
When adopting a UV-selected sample, the data presented in Figure 3 of \citet{Heinis14} is also consistent with
no redshift evolution in $f_{\mathrm{obscured}}$ from $z$$\sim$4 to $z$$\sim$1.5.

We have also compared to a subset of the \emph{Herschel}-selected sample of 
dusty star-forming galaxies (DSFG) in
the COSMOS field at $0.5<z<2.5$ presented in \citet{Casey14}.  
As this sample requires direct sub-mm detections, it will be inherently biased towards 
high fractions of obscured star formation, although the points are consistent within the errors.

Next, we compare our results to Equation 7 in \citet{Bethermin12}; this equation is the 
SFR$_{\mathrm{IR}}$/SFR$_{\mathrm{UV}}$ ratio based on the UV light attenuation measured in 
\citet{Pannella09}.  \citet{Pannella09} derived the UV and 1.4 GHz (which is assumed to be a proxy for IR) SFRs 
of a K-selected BzK sample in the COSMOS field, adopting photometric redshifts at $1<z<3$ for 
log(M$_{\star}$/M$_{\odot}$)$>$10.  We correct the stellar masses from the assumed Salpeter 
IMF to Chabrier IMF using the conversion of -0.24 dex in \citet{Zahid12}.  We convert this 
equation to the SFR$_{\mathrm{IR}}$/SFR$_{\mathrm{UV+IR}}$ ratio and include this relation 
in the left panel of Figure~\ref{fig:frac_obscured} for the stellar mass regime considered 
in Pannella et al. (dot-dash line). It is difficult to differentiate between the effects of 
sample selection (mass-selected vs. the BzK color-selection), data quality (grism vs. photometric 
redshifts), and SFR indicators (24$\mu$m vs. \emph{Herschel}). Despite these limitations, 
our result that $f_{\mathrm{obscured}}$ does not evolve with redshift agrees with the assumptions 
made in \citet{Bethermin12} based on data from \citet{Pannella09}.

Finally, we test in greater detail in Appendix A whether our assumption 
of the single log average of the DH02 
templates is driving the results presented in Figure~\ref{fig:frac_obscured}.  In other words,
are we sensitive to the common assumption of a redshift-independent IR spectral energy distribution (SED)?
\citet{Bethermin12} show evidence that the average IR SED of
star-forming galaxies evolves with redshift \citep[see also][]{Magdis12}. 
In the \citet{Bourne17} analysis, the models of \citet{Bethermin12} predict 
IR luminosities that are a factor of 1.2 higher on average at $2.5 < z < 4$.  We present a 
direct comparison of $f_{\mathrm{obscured}}$ when measured using the templates of 
\citet[][as used in Bethermin et al.]{Magdis12, Kirkpatrick12} and \citet{Kirkpatrick15}.  Although
we find a systematic offset towards lower values of $f_{\mathrm{obscured}}$ at log(M$_{\star}$/M$_{\odot}$)$<$10.5
than inferred using the DH02 templates, the overall trends remain redshift-independent.  
Redshift- and L$_{\mathrm{IR}}$-dependent template conversions to bolometric IR luminosity 
can significantly effect the measured $f_{\mathrm{obscured}}$ values but not the redshift evolution 
of the trend itself.

\begin{figure}[t]
\leavevmode
\centering
\includegraphics[width=\linewidth]{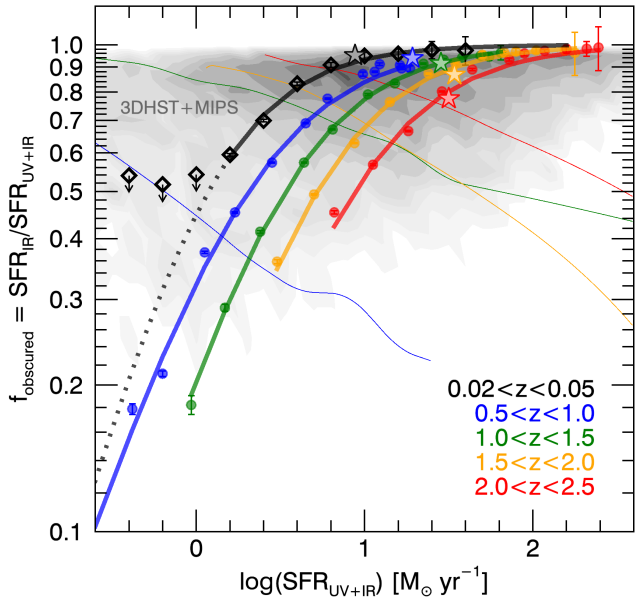}
\caption{The fraction of obscured star formation decreases with increasing redshift at fixed SFR$_{\mathrm{UV+IR}}$.  
The median values are derived from GALEX and WISE photometry
at 0.02$<$$z$$<$0.05 (black diamonds; Cybulski et al. in prep), and the 0.3--8$\mu$m 3D-HST photometry combined
with Spitzer/MIPS 24$\mu$m photometry at 0.5$<$$z$$<$2.5 \citep[circles;][]{Whitaker14b}.  The greyscale
shows the individual 3D-HST measurements at 0.5$<$$z$$<$2.5, with the thin solid lines representing the
respective completeness limits and thick-lines best-fit logistic growth functions. Star symbols represent
the inferred evolution of $f_{\mathrm{obscured}}$ of Milky-Way progenitors across cosmic time.}
\label{fig:frac_obscured_sfr}
\end{figure}

\subsection{How does the fraction of obscured star formation depend on total star formation rate?}
\label{sec:sfr}

\begin{table}[t]
\centering
\begin{threeparttable}
    \caption{Logistic Growth Function: $f_{\mathrm{obscured}}$ vs. log(SFR$_{\mathrm{UV+IR}}$)\label{tab:sfr_fit}}
    \begin{tabular}{lcc}
      \hline \hline
      & {$a$} & {$b$} \\
      \hline
      \noalign{\smallskip}
      $0.02<z<0.05$  &  1.234$\pm$0.033 & -2.858$\pm$0.083 \\
      $0.5<z<1.0$  &  2.092$\pm$0.011 & -2.384$\pm$0.012 \\
	  $1.0<z<1.5$ &  3.917$\pm$0.049 & -2.596$\pm$0.017 \\
      $1.5<z<2.0$  &  6.806$\pm$0.106 & -2.673$\pm$0.018 \\
       $2.0<z<2.5$ &  10.701$\pm$0.333 & -2.516$\pm$0.026 \\
      \noalign{\smallskip}
      \hline
      \noalign{\smallskip}
    \end{tabular}
    \begin{tablenotes}
      \small
    \item \emph{Notes.} Logistic growth function coefficients parameterizing the evolution of 
    	$f_{\mathrm{obscured}}$-$\log($SFR$_{\mathrm{UV+IR}}$) at different redshift epochs 
    	using Equation~\ref{eq:sfr_fit}.  
    \end{tablenotes}
  \end{threeparttable}
\end{table}

Next we explore how the fraction of obscured star formation depends 
on the total SFR$_{\mathrm{UV+IR}}$.  While we saw a marked lack of evolution in the median
trends with M$_{\star}$ out to $z$=2.5, we show in Figure~\ref{fig:frac_obscured_sfr} that
SFR$_{\mathrm{UV+IR}}$ increases dramatically.  However, the overall shape does not vary strongly.  
We include the median values from the SDSS $z\sim0$ sample (Cybulski et al. in prep); 
this data was reduced and analyzed completely independently, including a different
IR calibration (e.g., WISE 22$\mu$m vs. \emph{Spitzer}/MIPS 24$\mu$m), yet the evolution is consistent
over the full redshift range considered.  The best-fit relation for each redshift 
epoch is defined by a logistic growth function:

\begin{equation}
f_{\mathrm{obscured}}=\frac{1}{1+a e^{b\log(\mathrm{SFR_{UV+IR}})}}.
\label{eq:sfr_fit}
\end{equation}  

The strong redshift evolution of the median fraction of obscured star formation as a function of 
SFR$_{\mathrm{UV+IR}}$ can be understood generally by the increasing normalization of the star formation sequence,
as the global SFR is increasing towards earlier epochs in the Universe \citep{Madau14,Schreiber15,Tomczak16}.  
If we consider galaxies with the same fraction of obscured star formation,
we are effectively considering a fixed median stellar mass according to Figure~\ref{fig:frac_obscured}.  
Indeed, SFR$_{\mathrm{UV+IR}}$ increases by roughly 0.2 dex per $\Delta z\sim0.5$ at fixed $f_{\mathrm{obscured}}$.
For fractions $>$70\%, the evolution in SFR$_{\mathrm{UV+IR}}$ 
isn't as dramatic as the curves saturate at maximal levels of obscuration.
If we instead consider the evolution of $f_{\mathrm{obscured}}$ at fixed SFR, we will no longer 
be probing similar populations of star-forming
galaxies across cosmic time.  In this case, galaxies with the same SFR at high redshift
have significantly lower $f_{\mathrm{obscured}}$ than at low redshift.
\citet{Santini14} also measure lower dust mass per unit SFR at higher redshifts, 
in agreement with the results herein.

How do galaxies evolve in this diagram? In reality, we know that galaxies grow with time 
and therefore fixing stellar mass doesn't ensure
we are tracking similar galaxies across cosmic time.  To build some intuition, we can instead select
galaxies based on fixed number density and use Equation 1 of \citet{vanDokkum13} to estimate the redshift evolution 
of stellar mass for Milky-way progenitors, M$_{\mathrm{MW}}$.  When combining M$_{\mathrm{MW}}$($z$) 
with $f_{\mathrm{obscured}}$(M$_{\star}$) from Equation~\ref{eq:mass_fit} in this paper, we can 
predict $f_{\mathrm{obscured}}(z)$.  By adopting the SFR implied by the log(SFR)-log(M$_{\star}$) relation 
at a given $z$ \citep[see appendix in][]{Whitaker17}, we can predict the trajectory of a Milky-Way progenitor
in $f_{\mathrm{obscured}}$ and log(SFR) (star symbols in Figure~\ref{fig:frac_obscured_sfr}).  In this case,
$f_{\mathrm{obscured}}$ changes more rapidly at higher redshift, increasing by 14\% from $z$=2.25 to $z$=1.25 (2 Gyr),
but only 4\% from $z$=1.25 to $z$=0 (8 Gyr).

\section{Discussion}
\label{sec:discussion}

In this paper, we demonstrate that at any given fraction of obscured star formation 
($f_{\mathrm{obscured}}$=SFR$_{\mathrm{IR}}$/SFR$_{\mathrm{UV+IR}}$), there is little evolution in the median 
trends with stellar mass over 11 billion years of cosmic time ($z$=0 to $z$=2.5).
Interestingly, studies have also found little to no redshift evolution when considering 
A$_{1500}$ \citep[e.g.,][]{Pannella09}, A$_{\mathrm{H\alpha}}$ \citep[e.g.,][]{Sobral12}, 
and A$_{\mathrm{V}}$ \citep[e.g.,][]{Martis16} at fixed stellar mass, all quantities parameterizing the 
amount of dust attenuation.
This marks the very epoch over which the SFR density peaks and drops precipitously \citep{Madau14}, 
and the metallicities of galaxies of a given mass change by $\sim$0.2--0.6 
dex \citep[e.g.,][]{Savaglio05, Erb06a, Kewley08}, with more evolution at the low stellar mass end. 
Galaxies at higher redshift have lower metallicities and therefore one would expect them to produce less dust. 
Consequently, if there is a direct scaling between dust mass and L$_{\mathrm{IR}}$, 
less star formation would be obscured in these galaxies relative to similar masses at low redshift.
It is therefore not immediately clear why there is a lack of redshift evolution 
in $f_{\mathrm{obscured}}$ (and dust attenuation) at fixed stellar mass.

\begin{figure*}[t]
\leavevmode
\centering
\includegraphics[width=0.78\linewidth]{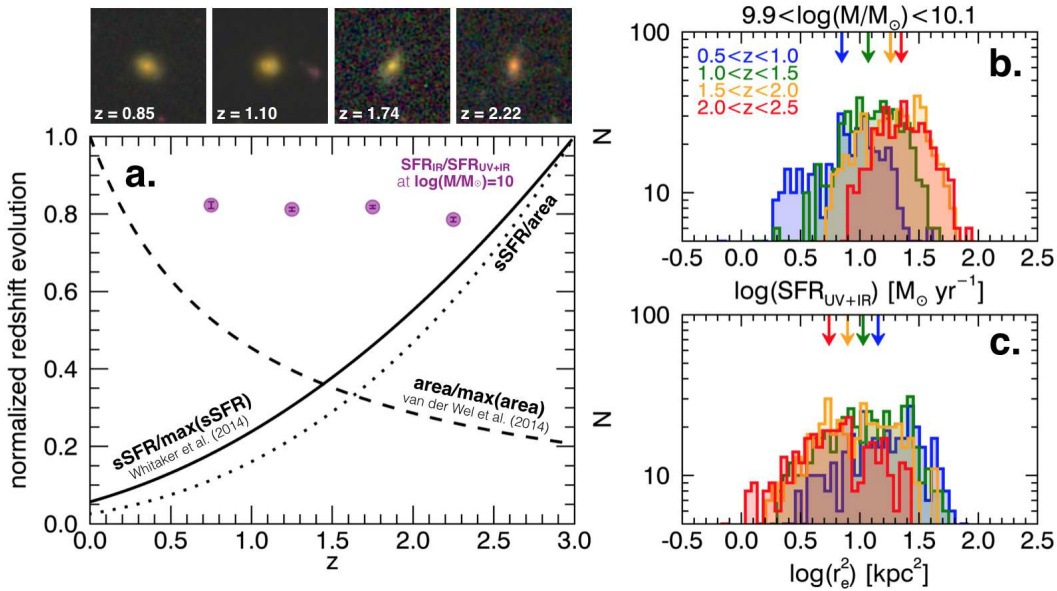}
\caption{Galaxies selected in a narrow range of stellar mass (e.g., 9.9$<$log(M$_{\star}$/M$_{\odot}$)$<$10.1) 
exhibit similar values of $f_{\mathrm{obscured}}$ (e.g., $\sim$80\%) out to $z$=2.5 (purple circles in panel, taken from 
Figure~\ref{fig:frac_obscured}).  
Yet the average size and SFR properties are vastly different at different epochs: 
$r_{e}^{2}$ decreases with increasing redshift (dashed line), 
while the average sSFR (solid line) and sSFR/area (dotted line) both increase.
Given the narrow range in stellar mass, our selection is equivalent to the evolution of 
SFR and SFR/area (or SFR surface density). 
The mean of the distributions of log(SFR) (panel b) and log($r_{e}^{2}$) 
(panel c) demonstrate similar opposing redshift evolution, with arrows marking the median.  
Even though galaxies 
exhibit significantly different SFR and size properties at different redshifts on average, as also demonstrated with the 
example three-color \emph{HST} postage stamps, there is little redshift evolution of their dust properties on a whole.}
\label{fig:trends}
\end{figure*}

The amount of dust in these moderate to massive galaxies ($>$10$^{9}$ M$_{\odot}$) is 
effectively set by the balance between dust production and destruction.  Dust production
depends on the growth in the interstellar medium (ISM) via the process of coagulation within molecular clouds.
Key additional production channels including supernovae (SNe) and stellar ejecta from stars in the asymptotic giant 
branch phase of stellar evolution. The ionizing UV and X-ray radiation from massive stars can also easily
destroy dust grains, in addition to collisional destruction.
While the ISM growth is governed to first order by the volume density of gas 
and the metallicity, SNe and stellar winds are affected by the SFR
volume density \citep[e.g.,][]{Popping17}.
Given that the analysis herein is empirically driven in nature, we next explore the 
how the effects of dust production and destruction could explain the lack of redshift evolution
of the $f_{\mathrm{obscured}}$-log(M$_{\star}$) relation predominantly 
in the context of existing observational results.

\subsection{Metallicity}

First, we will consider the expected effects of metallicity on the amount of dust production through
correlations with both dust-to-gas ratio and stellar mass.
Observations of local galaxies show that dust-to-gas ratio scales linearly with metallicity above 
$\sim$0.1 Z$_{\odot}$ \citep[e.g.,][]{Draine07, Leroy11, Remy14}. This marks a reasonable metallicity
threshold for the present study, given that the vast majority of galaxies in our sample should 
have Z$>$0.1Z$_{\odot}$.  
If we assume the dust-to-gas ratio does not vary strongly with redshift, as suggested by models 
\citep{Feldmann15, Popping17}, we can calculate how this quantity scales with stellar mass at 
various redshift epochs.  We describe the details of this analysis in Appendix \ref{sec:appendixB}.
Our compilation of observations of mass-metallicity and dust-to-gas ratio vs. metallicity 
suggest that the dust-to-gas ratio is a factor of $\sim$3 larger 
at log(M/M$_{\odot}$)=11 as compared to log(M/M$_{\odot}$)=9, with minimal redshift evolution when 
normalized to the dust-to-gas ratio for the most massive galaxies at a given epoch.  
In Figure~\ref{fig:frac_obscured}, we see a similar relative difference in the fraction of obscured 
star formation for the same stellar mass range. However, the overall shapes don't quite match 
in detail; the increase at lower stellar masses is more dramatic for the fraction of obscured 
star formation relative to change in dust-to-gas ratio, with a 30\% larger increase from log(M/M$_{\odot}$)=9--10
(see Figure~\ref{fig:appendixB}).

The tension in the comparison of the fraction of obscured star formation to dust-to-gas ratios as 
a function of stellar mass will only be amplified when accounting for the fact that gas fractions 
decrease with stellar mass at a given 
epoch \citep[e.g.][]{Narayanan15, Morokuma15, Popping15, Scoville16, Saintonge11, Saintonge16, Tacconi13, Tacconi17}. 
Owing to the increased gas fraction of lower mass galaxies relative to more massive galaxies at a given 
redshift, the relation between dust mass to stellar mass will be shallower than that with dust-to-gas ratio.
The total amount of dust (or the column of dust) seen by the UV photons 
will therefore be higher for lower mass galaxies relative to the dust-to-gas ratio trend, 
suggesting a higher SFR$_{\mathrm{IR}}$ and hence SFR$_{\mathrm{IR}}$/SFR$_{\mathrm{UV+IR}}$.  
Although this could in principle increase the fraction
of obscured star formation at the low mass end, it acts in the wrong direction to 
alleviate the discrepancies.  
We therefore find that the drop in $f_{\mathrm{obscured}}$ we observe at 
lower M$_{\star}$ appears to be a stronger function of stellar mass than explainable by the stellar
mass dependence of the dust-to-gas ratio.

\subsection{Surface Density}

Next, we consider the effects of gas surface density on the amount of dust production.
An increase in the gas surface density will correlate with an increase in the SFR surface 
density \citep{Schmidt59, Kennicutt98b}, as well as the column density of dust.  
Consequently, any redshift evolution in the gas/SFR surface densities would suggest an 
increase in the level of dust-obscured star formation.  We next consider implications from 
the well-studied average scaling relations of $r_{e}$ and SFR that comprise the SFR surface density.
At fixed stellar mass, observations show that while the average galaxy size (and area) 
decreases \citep[e.g.,][]{vanderWel14}, the average SFR increases with 
redshift \citep[e.g.,][]{Whitaker14b}.  The SFR surface density 
of galaxies therefore increases on average towards higher redshift \citep[e.g.,][]{Wisnioski12, Livermore15}.
This is corroborated by a recent archival study by \citet{Fujimoto17} of ALMA/1mm images, who
find that the average size of the gas reservoirs follows a similar redshift evolution to that
of the rest-frame optical sizes measured from \emph{HST} imaging.  However, the gas size is statistically
smaller, suggesting that dust-obscured star formation occurs in compact regions.
When combining these various empirical trends with observations of increasing gas fractions 
\citep[e.g.,][]{Tacconi17}, the data suggest higher gas column densities that produce 
conditions more suitable to form dust at high redshift,
both through star formation and the accretion of metals.

Yet, it may be that the average redshift evolution of the SFR density and galaxy size
cancel each other out such that $f_{\mathrm{obscured}}$ does not evolve with redshift for a given M$_{\star}$. 
We show the normalized average redshift evolution in 
area ($r_{e}^{2}$) and (s)SFR$_{\mathrm{UV+IR}}$ for a narrow range of 
9.9$<$log(M$_{\star}$/M$_{\odot}$)$<$10.1 in
Figure~\ref{fig:trends} (panel a).  The narrow mass bin is intentional such that we select a sample of 
galaxies that have a similar $f_{\mathrm{obscured}}$ on average.
We also show the trend of $f_{\mathrm{obscured}}$ with redshift in the equivalent stellar mass bin (purple).
Note that we are not calculating sSFR/r$_{e}^{2}$ for any particular individual 
galaxy but rather using the independently measured trends between SFR and r$_{e}$ with stellar mass.
Given that increased gas and SFR surface densities imply higher dust production 
at earlier times, it follows that more compact galaxy 
geometries would decrease dust creation if we want to explain the
observed constant median $f_{\mathrm{obscured}}$ per M$_{\star}$ with redshift (purple). 
This is counter-intuitive and poses a challenge 
to current dust models. Possible explanations can include elevated 
dust destruction resulting from compact geometries and increased optical thickness.  
Denser star-forming environments will have an elevated number of supernovae explosions 
on relatively short timescales, which can either destroy or remove gas and dust in outflows. 
In the case of more massive galaxies, self-absorption due to the increased optical 
thickness from higher gas fractions may also play a role. 

We see in the right panels of Figure~\ref{fig:trends} that while the median changes with redshift, 
the distributions of both sSFR and r$_{e}^{2}$ largely overlap with one another. 
After removing the well known correlations between stellar mass and SFR and r$_{e}$, 
\citet{Whitaker17} demonstrate that star-forming galaxies show only a weak dependence of SFR on r$_{e}$.  
As there is little to no correlation between r$_{e}^{2}$ and (s)SFR,  
the scatter in $f_{\mathrm{obscured}}$ for individual galaxies of a similar stellar mass may be quite 
large in reality.  Stacking could hide weaker trends that exist within the scatter, especially if the trends 
are stronger for some subset of the galaxies, for example. 
Drawing connections between average galaxy correlations and underlying 
dust physics is not trivial.  Figure~\ref{fig:trends} serves to caution the reader that  
intrinsic scatter in the physical properties of galaxies may result in a more complicated 
redshift evolution of $f_{\mathrm{obscured}}$ on a galaxy by galaxy basis.

\subsection{Uncertainties in Dust Creation and Destruction}

It is worth noting that the obscuration of UV photons is far more complicated than
our simplified assumptions regarding the dependence on metallicity or dust column density.  
Theoretical models must also 
explicitly take into account how the dust mass, dust-to-gas ratios, and dust column densities translate
into the fraction of obscured star formation.  It has been shown that the global IR 
luminosity is not necessarily directly correlated with total dust mass, and consequently SFR$_{\mathrm{IR}}$, 
as the dust grains in different physical regions are subject to a range of radiation field strengths and hence emit
over a range of blackbody temperatures in reality \citep[e.g.][]{Dunne00,Draine07,Magdis12,Kirkpatrick17}.
Moreover, dust geometry can change the extinction
law in dense environments around nascent stars, either via dust growth by accretion, grain growth
by coagulation where small dust grains stick to larger ones, or grain disruption by shattering 
where larger grains are shattered by collisions \citep[e.g.,][]{Hirashita12, Hirashita15}.  
Such processes will change the UV absorption properties of the ISM. And, more generally, empirical 
studies show that the best-fit attenuation law varies with galaxy type and physical 
properties \citep[e.g.,][]{Wild11, Kriek13, Reddy15, Zeimann15, Salmon16, Battisti16}.
The time evolution of these processes together with the observational uncertainties in the 
empirical trends quickly results in a complicated picture.

It also may be that global measures of IRX (analogous SFR$_{\mathrm{IR}}$
to SFR$_{\mathrm{UV}}$) and SFR$_{\mathrm{IR}}$/SFR$_{\mathrm{UV+IR}}$ are imperfect tracers of 
true obscuration within galaxies. 
Are star-forming regions cohabitants with the majority of dust in the ISM?  Several case studies 
find significant offsets between sub-mm and rest-UV/optical emission \citep[e.g.,][]{Iono06,Chen15,Koprowski16}, 
suggesting that the bulk of the UV and IR emission may originate from different physical regions 
in star-forming galaxies at high redshift. 
If this is a ubiquitous feature of galaxies, the interpretation of a global ratio comparing UV and IR
emission becomes muddled. Though beyond the scope of this work, future spatially resolved analyses
will illuminate the geometric effects at play.

Though disentangling the complex interplay between dust geometry and composition with other
physical properties of galaxies is beyond the
scope of this empirical work, the non-evolution in redshift of the median $f_{\mathrm{obscured}}$ 
as a function of M$_{\star}$ points to very little redshift evolution in the characteristics of dust on a whole. 
Rigorous theoretical analyses will be required
to understand the exact balance between the dust creation and destruction processes mentioned
above across time in the context of these empirical results.

\section{Conclusions}
\label{sec:conclusions}

In this paper, we explore how the total star formation rate and stellar masses of galaxies depend on the
relative amount of obscured star formation.  Our main mass-complete galaxy sample is comprised of 39,106 star-forming galaxies 
at $0.5<z<2.5$ \citep{Whitaker14b}, selected from the 3D-HST/CANDELS treasury programs
in the five premier extragalactic fields.  This deep near-infrared
photometry yields mass-complete galaxy samples down to unprecedented limits of log(M/M$_{\odot}$) = 8.7 (9.3) at $z=1.0$ ($z=2.5$).
Unobscured star formation rates are measured directly from the 3D-HST photometry, and obscured star formation
is quantified based on stacks of deep \emph{Spitzer}/MIPS 24$\mu$m photometry. 
We expand the baseline of the analysis by combining this novel data set with a local SDSS sample of 22,481 star-forming galaxies at 
0.02$<$$z$$<$0.05, with total star formation rates measured from GALEX and WISE photometry.

The main findings of our analysis are summarized as:
\vspace{-0.3cm}

\begin{enumerate}

	\item We observe a strong dependence of the median fraction of obscured star formation 
	(defined as $f_{\mathrm{obscured}}$=SFR$_{\mathrm{IR}}$/SFR$_{\mathrm{UV+IR}}$) 
    on stellar mass.  This correlation shows remarkably little evolution across the full 
    redshift range explored ($z=0$ to $z=2.5$), extending earlier results 
    \citep[e.g.,][]{Heinis14,Bourne17} to lower stellar mass limits.
	\vspace{-0.3cm}
	\item The transition from mostly unobscured to obscured star formation ($f_{\mathrm{obscured}}$=0.5) occurs
	at a relatively low stellar mass of log(M/M$_{\odot}$)=9.4.  Even though the majority of the star formation in galaxies with 
	log(M/M$_{\odot}$)$<$9.4 is radiated in the rest-frame UV, there exists a tail of low-mass extremely obscured star-forming
	galaxies at $z>1$.  For the most massive galaxies, with log(M/M$_{\odot}$)$>$10.5, $>$90\% of star formation is obscured 
	at all redshifts.  
	\vspace{-0.3cm}	
	\item We find that the fraction of star formation obscured by dust,$f_{\mathrm{obscured}}$, at fixed total 
	SFR decreases at higher redshift. As the normalization of the log(SFR)-log(M$_{\star}$) relation is increasing with redshift
	while $f_{\mathrm{obscured}}$-log(M$_{\star}$) is unchanged, the same total SFR probes lower stellar mass limits 
	at higher redshifts.
\end{enumerate}

We explore the implications of these findings in the context of a range of well-studied empirical trends between
dust-to-gas ratio, metallicity, gas and SFR surface density, gas fraction and stellar mass. 
Galaxies at high redshift are observed to have more compact sizes and significantly higher star formation rates,
gas fractions, and hence gas and SFR surface densities.  The straight-forward interpretation is 
therefore that star formation should be more highly obscured at early times.  It is therefore
puzzling that we observe no redshift evolution in the median fraction of obscured star formation with stellar
mass out to $z=2.5$.  Given the complexity of the various physical processes governing the attenuation
of UV photons, including but not limited to dust geometry and composition, key progress can be made in understanding
these results with future theoretical models.

\begin{acknowledgements}
We thank the anonymous referee for useful comments and a careful reading of the paper.
The authors wish to acknowledge P. van Dokkum, I. Momcheva, R. Skelton, G. Brammer, the 3D-HST
team and colleagues for their hard work in releasing public data and 
catalogs in the 3D-HST fields.  KEW is grateful for discussions with N. Katz.
KEW gratefully acknowledges support by NASA through Hubble Fellowship grant 
\#HST-HF2-51368 awarded by the Space Telescope Science Institute, which is operated 
by the Association of Universities for Research in Astronomy, Inc., for NASA, under 
contract NAS 5-26555. 
MSY and RC acknowledges support from the NASA ADAP grant NNX14AF80G.
CMC thanks the UT Austin College of Natural Science for support.
This work is based on observations taken by the 3D-HST Treasury Program (GO 12177 and 12328)
with the NASA/ESA HST, which is operated by the Associations of Universities for Research in Astronomy, Inc.,
under NASA contract NAS5-26555.
\end{acknowledgements}

\appendix
\renewcommand{\thefigure}{A\arabic{figure}}
\setcounter{figure}{0}

\section{Appendix A.}
\label{sec:appendixA}

\begin{figure*}[b]
\leavevmode
\centering
\includegraphics[width=0.85\linewidth]{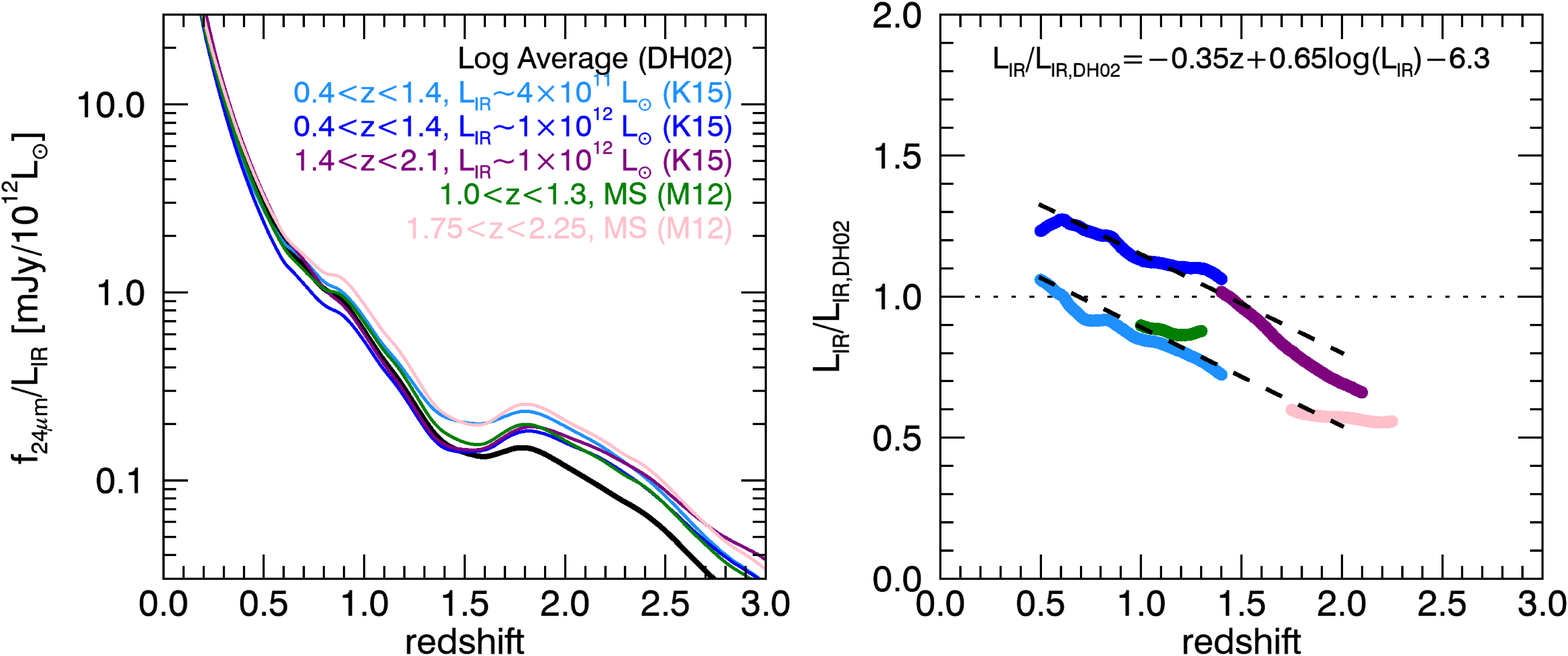}
\caption{(Left) Comparison of different template conversions transforming 24$\mu$m flux 
density to L$_{\mathrm{IR}}$ relative to the redshift-independent DH02 templates adopted herein.  We compare to 
the empirical templates of \citet{Kirkpatrick15} (K15) and \citet{Magdis12} (M12).  
The K15 templates explore a range of L$_{\mathrm{IR}}$ for two redshift intervals,
whereas the two M12 templates shown are for typical star-forming galaxies at different redshift epochs. (Right)
The resulting L$_{\mathrm{IR}}$ from the K15 and M12 templates relative to the DH02 templates mildly
evolves with redshift.  We parameterize this evolution as a function of redshift and L$_{\mathrm{IR}}$ (dashed lines) to test how this affects the obscured SFR presented in this paper.}
\label{fig:templates}
\end{figure*}

In order to explore the effect of template uncertainties on the calculation of IR SFRs, we compare 
our data to the template set adopted within \citet{Bethermin12} and \citet[][hereafter K15]{Kirkpatrick15}.  
Bethermin et al. use a redshift-dependent IR SED template set for ``typical'' star-forming galaxies 
and starbursts from \citet[][hereafter M12]{Magdis12}, based on fits of the \citet{Draine07} models.  
K15 published empirical composite templates of star-forming galaxies at $0.4<z<1.4$ for L$_{\mathrm{IR}}\sim4\times10^{11}$ 
and $1\times10^{12}$ L$_{\odot}$, plus a higher redshift template at $1.4<z<2.1$ for 
L$_{\mathrm{IR}}\sim1\times10^{12}$ L$_{\odot}$.
We use the K15 together with the M12 templates for ``main sequence'' star-forming galaxies at $1.0<z<1.3$ and $1.75<z<2.25$ to 
recalculate L$_{\mathrm{IR}}$ for all the galaxies in the 3D-HST catalog that have a positive 24$\mu$m 
flux densities at the relevant redshifts (Figure~\ref{fig:templates}).   

\begin{figure*}[b]
\leavevmode
\centering
\includegraphics[width=\linewidth]{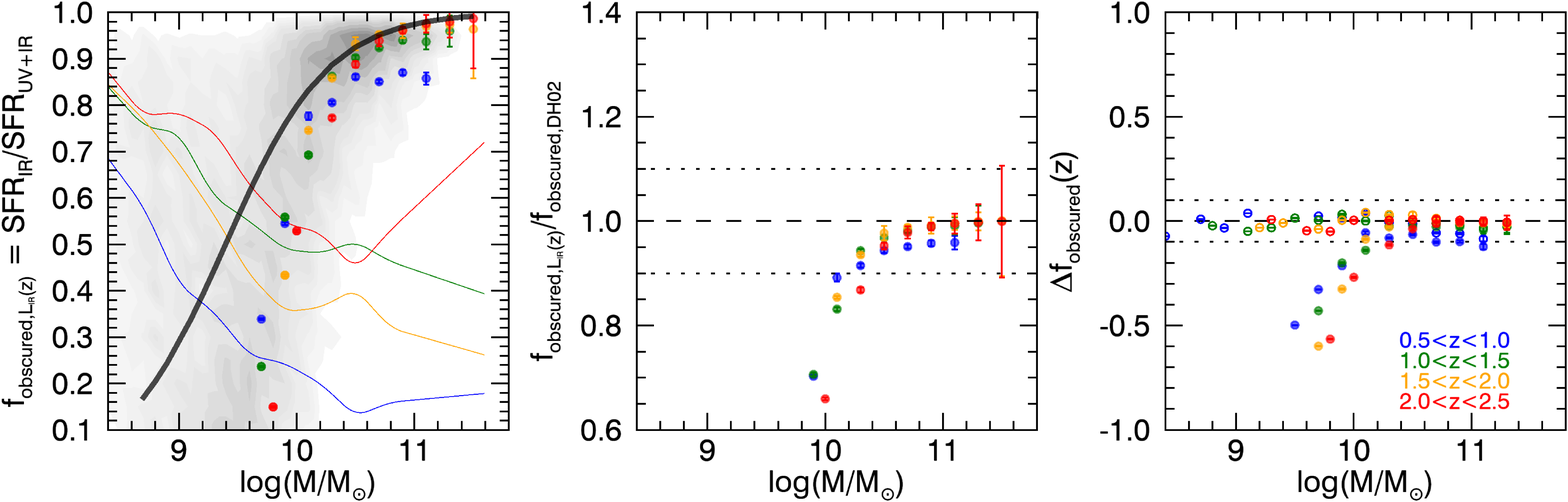}
\caption{(Left) Changing the template conversion of 24$\mu$m flux density to L$_{\mathrm{IR}}$, using the equation
in Figure~\ref{fig:templates}b, changes the overall shape of $f_{\mathrm{obscured}}$ as a function of 
stellar mass, but the relation still does not vary strongly with redshift. This is demonstrated in the middle and
right panels, where we compare the ratio of the new $f_{\mathrm{obscured}}$ relative to the original using the
DH02 templates (middle panel) and the difference between the new $f_{\mathrm{obscured}}$ and the best-fit model (right panel).  
The scatter between the original data and best-fit line (open circles) increases slightly when considering the K15/M12
template conversions (filled circles). The difference due to template conversion is negligible at log(M/M$_{\odot}$)$>$10.5.}  
\label{fig:frac_obscured_templates}
\end{figure*}

The right panel in Figure~\ref{fig:templates} shows the ratio of 
L$_{\mathrm{IR}}$ from the K15 and M12 templates to the DH02 template used in \citet{Whitaker12b, Whitaker14b} and this paper, 
color-coded based on the template adopted.  In general, we find that the M12 and K15 templates predict a 
mild redshift and L$_{\mathrm{IR}}$ dependence on the conversion from 24$\mu$m flux density to bolometric IR luminosity 
(Figure~\ref{fig:templates}, right panel). The trend is consistent between the K15 and M12 templates, 
though M12 does not explicitly isolate galaxy samples based on L$_{\mathrm{IR}}$.
The dashed lines in the right panel show the best-fit linear function, 
which we quantify as a function of redshift and L$_{\mathrm{IR}}$ to be:  

\begin{equation}
L_{\mathrm{IR}}(z)/L_{\mathrm{IR,DH02}} = -0.35z + 0.65\times\log(L_{\mathrm{IR}}) - 6.3
\label{eq:templates}
\end{equation}

In order to isolate if template-dependent effects are driving the results shown in Figure~\ref{fig:frac_obscured}, 
we correct L$_{\mathrm{IR}}$ in the 3D-HST catalogs using the equation above and recalculate 
SFR$_{\mathrm{IR}}$/SFR$_{\mathrm{UV+IR}}$ vs. log(M$_{\star}$). To avoid over-interpreting 
the K15 and M12 templates, we only consider the regime of log(M$_{\star}$/M$_{\odot}$)$>$9.7, 
which corresponds to the lower mass limit of galaxies in the M12 sample.  Although the K15 sample 
does not include such low mass galaxies, the results are consistent between the two independent 
template sets (Right panel, Figure~\ref{fig:templates}). The ratio of the measured values of 
$f_{\mathrm{obscured}}$ = SFR$_{\mathrm{IR}}$/SFR$_{\mathrm{UV+IR}}$ from the redshift- and 
L$_{\mathrm{IR}}$-dependent relation relative to the original redshift-independent DH02 template 
decreases from $\sim$30\% (0.15 dex) difference at log(M$_{\star}$/M$_{\odot}$)$\sim$10 to $<$5\% ($<$0.02 dex) difference
at log(M$_{\star}$/M$_{\odot}$)$>$10.5.  

In Figure~\ref{fig:frac_obscured_templates}, we compare $f_{\mathrm{obscured}}$ corrected for the 
redshift- and L$_{\mathrm{IR}}$-dependence of the templates relative to the original stacks based on 
the DH02 templates.  The middle panel directly compares to the original data, 
whereas the right panel is comparing the corrected $f_{\mathrm{obscured}}$ to the redshift-independent 
best-fit to the original data (shown as the thick black line in left-most panel for reference).  
The M12 and K15 empirical templates suggest that L$_{\mathrm{IR}}$ is over-predicted for 
lower mass galaxies with log(M$_{\star}$/M$_{\odot}$)$\sim$10, such that $f_{\mathrm{obscured}}$ 
is also overpredicted by as much as 50\% (0.3 dex).  However, $f_{\mathrm{obscured}}$ agrees 
regardless of the template conversion adopted for more massive galaxies above log(M$_{\star}$/M$_{\odot}$)$>$10.5 
within $<$5\% (0.02 dex). Interestingly, the analysis presented in \citet{Wuyts11a} found that while
L$_{\mathrm{IR}}$ derived using 24$\mu$m flux densities with the full DH02 template set is over-predicted 
for the most massive, higher redshift galaxies (log(SFR)$>$2 M$_{\odot}$ yr$^{-1}$) in our sample 
by up to $\sim$0.5 dex, the systematic offset is much weaker for the log-average of the DH02 templates.  
We show here that using an evolving template set for the 24$\mu$m conversion 
may alleviate any remaining discrepancy. 

Despite different trends that emerge when employing a different template conversion from the 24$\mu$ 
flux density to bolometric IR luminosity, the lack of redshift evolution in $f_{\mathrm{obscured}}$ 
with stellar mass remains a robust conclusion. Namely, the scatter between $f_{\mathrm{obscured}}$ for different 
redshift epochs in the middle and right-most panel of Figure~\ref{fig:frac_obscured_templates} 
remains small, even when utilizing a redshift dependent conversion. 

\section{Appendix B.}
\label{sec:appendixB}

\renewcommand{\thefigure}{B\arabic{figure}}
\setcounter{figure}{0}

\begin{figure}
\leavevmode
\centering
\includegraphics[width=0.45\linewidth]{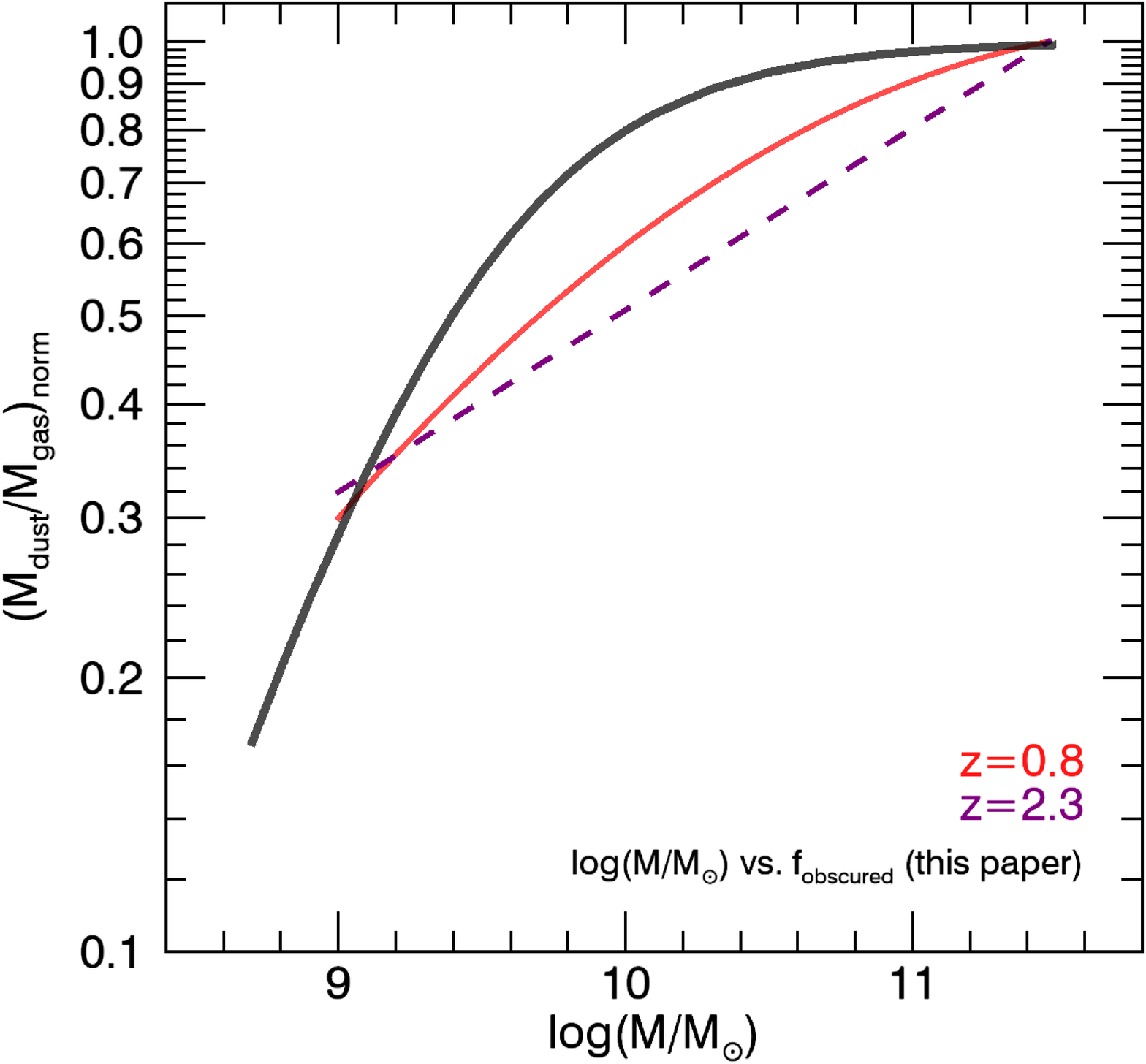}
\caption{The relative change in the ratio of dust-to-gas mass with stellar mass (normalized to unity at log(M/M$_{\odot}$)=11.5)
at $z$=0.8 (purple) and $z$=2.3 (red) is similar to that of $f_{\mathrm{obscured}}$:  Massive galaxies with log(M/M$_{\odot}$)=11 
have a factor of three higher dust-to-gas mass and $f_{\mathrm{obscured}}$ than lower mass galaxies, log(M/M$_{\odot}$)=9.  
However, the overall shape of the dust-to-gas mass as a function of stellar mass 
relative to the stellar mass dependence of $f_{\mathrm{obscured}}$ (black, from Figure~\ref{fig:frac_obscured}) is different.}
\label{fig:appendixB}
\end{figure}

In order to directly compare the dust-to-gas ratio as a function of stellar mass, we take two
well measured correlations:  the dust-to-gas ratio as a function of metallicity  from \citet{Remy14},
and mass-metallicity relations that bookend our redshift distribution.
We adopt the mass-metallicity relation presented in \citet{Zahid11} 
at $z$$\sim$0.8 from the AEGIS/DEEP-2 spectroscopic sample, and \citet{Steidel14} at $z$$\sim$2.3
from KBSS-MOSFIRE spectroscopy.  The mass-metallicity relation results at $z$$\sim$2.3 from the MOSDEF survey 
\citep{Sanders15} are consistent with \citet{Steidel14} when selecting targets from the 
mass-complete 3D-HST parent sample.  In Figure~\ref{fig:appendixB}, we normalize 
the parameterized relation between dust-to-gas ratio and stellar mass at $z$=0.8 (purple) 
and $z$=2.3 (red) at log(M/M$_{\odot}$)=11.5 to facilitate a direct comparison
with the best-fit relation from Figure~\ref{fig:frac_obscured} (black).  We explore the relative
dependence of dust-to-gas ratio and $f_{\mathrm{obscured}}$ on stellar mass in the main
body of the paper.

\addcontentsline{toc}{chapter}{\numberline {}{\sc References}}


\begin{thebibliography}{96}
\expandafter\ifx\csname natexlab\endcsname\relax\def\natexlab#1{#1}\fi

\bibitem[{{Alam} {et~al.}(2015){Alam}, {Albareti}, {Allende Prieto}, {Anders},
  {Anderson}, {Anderton}, {Andrews}, {Armengaud}, {Aubourg}, {Bailey}, \&
  et~al.}]{Alam15}
{Alam}, S., {Albareti}, F.~D., {Allende Prieto}, C., {et~al.} 2015, \apjs, 219,
  12

\bibitem[{{Battisti} {et~al.}(2016){Battisti}, {Calzetti}, \&
  {Chary}}]{Battisti16}
{Battisti}, A.~J., {Calzetti}, D., \& {Chary}, R.-R. 2016, \apj, 818, 13

\bibitem[{{Bell} {et~al.}(2005){Bell}, {Papovich}, {Wolf}, {Le Floc'h},
  {Caldwell}, {Barden}, {Egami}, {McIntosh}, {Meisenheimer},
  {P{\'e}rez-Gonz{\'a}lez}, {Rieke}, {Rieke}, {Rigby}, \& {Rix}}]{Bell05}
{Bell}, E.~F., {Papovich}, C., {Wolf}, C., {et~al.} 2005, \apj, 625, 23

\bibitem[{{Bertin} \& {Arnouts}(1996)}]{Bertin96}
{Bertin}, E., \& {Arnouts}, S. 1996, \aaps, 117, 393

\bibitem[{{B{\'e}thermin} {et~al.}(2012){B{\'e}thermin}, {Daddi}, {Magdis},
  {Sargent}, {Hezaveh}, {Elbaz}, {Le Borgne}, {Mullaney}, {Pannella}, {Buat},
  {Charmandaris}, {Lagache}, \& {Scott}}]{Bethermin12}
{B{\'e}thermin}, M., {Daddi}, E., {Magdis}, G., {et~al.} 2012, \apjl, 757, L23

\bibitem[{{Bourne} {et~al.}(2017){Bourne}, {Dunlop}, {Merlin}, {Parsa},
  {Schreiber}, {Castellano}, {Conselice}, {Coppin}, {Farrah}, {Fontana},
  {Geach}, {Halpern}, {Knudsen}, {Micha{\l}owski}, {Mortlock}, {Santini},
  {Scott}, {Shu}, {Simpson}, {Simpson}, {Smith}, \& {van der Werf}}]{Bourne17}
{Bourne}, N., {Dunlop}, J.~S., {Merlin}, E., {et~al.} 2017, \mnras, 467, 1360

\bibitem[{{Brammer} {et~al.}(2008){Brammer}, {van Dokkum}, \&
  {Coppi}}]{Brammer08}
{Brammer}, G.~B., {van Dokkum}, P.~G., \& {Coppi}, P. 2008, \apj, 686, 1503

\bibitem[{{Brammer} {et~al.}(2011){Brammer}, {Whitaker}, {van Dokkum},
  {Marchesini}, {Franx}, {Kriek}, {Labb{\'e}}, {Lee}, {Muzzin}, {Quadri},
  {Rudnick}, \& {Williams}}]{Brammer11}
{Brammer}, G.~B., {Whitaker}, K.~E., {van Dokkum}, P.~G., {et~al.} 2011, \apj,
  739, 24

\bibitem[{{Bruzual} \& {Charlot}(2003)}]{BC03}
{Bruzual}, G., \& {Charlot}, S. 2003, \mnras, 344, 1000

\bibitem[{{Calzetti}(2013)}]{Calzetti13}
{Calzetti}, D. 2013, {Star Formation Rate Indicators}, ed.
  J.~{Falc{\'o}n-Barroso} \& J.~H. {Knapen}, 419

\bibitem[{{Calzetti} {et~al.}(2000){Calzetti}, {Armus}, {Bohlin}, {Kinney},
  {Koornneef}, \& {Storchi-Bergmann}}]{Calzetti00}
{Calzetti}, D., {Armus}, L., {Bohlin}, R.~C., {et~al.} 2000, \apj, 533, 682

\bibitem[{{Casey} {et~al.}(2014{\natexlab{a}}){Casey}, {Narayanan}, \&
  {Cooray}}]{Casey14a}
{Casey}, C.~M., {Narayanan}, D., \& {Cooray}, A. 2014{\natexlab{a}}, \physrep,
  541, 45

\bibitem[{{Casey} {et~al.}(2014{\natexlab{b}}){Casey}, {Scoville}, {Sanders},
  {Lee}, {Cooray}, {Finkelstein}, {Capak}, {Conley}, {De Zotti}, {Farrah},
  {Fu}, {Le Floc'h}, {Ilbert}, {Ivison}, \& {Takeuchi}}]{Casey14}
{Casey}, C.~M., {Scoville}, N.~Z., {Sanders}, D.~B., {et~al.}
  2014{\natexlab{b}}, \apj, 796, 95

\bibitem[{{Chabrier}(2003)}]{Chabrier}
{Chabrier}, G. 2003, \pasp, 115, 763

\bibitem[{{Chen} {et~al.}(2015){Chen}, {Smail}, {Swinbank}, {Simpson}, {Ma},
  {Alexander}, {Biggs}, {Brandt}, {Chapman}, {Coppin}, {Danielson},
  {Dannerbauer}, {Edge}, {Greve}, {Ivison}, {Karim}, {Menten}, {Schinnerer},
  {Walter}, {Wardlow}, {Wei{\ss}}, \& {van der Werf}}]{Chen15}
{Chen}, C.-C., {Smail}, I., {Swinbank}, A.~M., {et~al.} 2015, \apj, 799, 194

\bibitem[{{Cortese} {et~al.}(2006){Cortese}, {Boselli}, {Buat}, {Gavazzi},
  {Boissier}, {Gil de Paz}, {Seibert}, {Madore}, \& {Martin}}]{Cortese06}
{Cortese}, L., {Boselli}, A., {Buat}, V., {et~al.} 2006, \apj, 637, 242

\bibitem[{{Dale} \& {Helou}(2002)}]{DH02}
{Dale}, D.~A., \& {Helou}, G. 2002, \apj, 576, 159

\bibitem[{{Dickinson} \& {FIDEL Team}(2007)}]{Dickinson07}
{Dickinson}, M., \& {FIDEL Team}. 2007, in Bulletin of the American
  Astronomical Society, Vol.~39, American Astronomical Society Meeting
  Abstracts, 822

\bibitem[{{Dickinson} {et~al.}(2003){Dickinson}, {Papovich}, {Ferguson}, \&
  {Budav{\'a}ri}}]{Dickinson03}
{Dickinson}, M., {Papovich}, C., {Ferguson}, H.~C., {et~al.} 2003, \apj, 587,
  25

\bibitem[{{Donley} {et~al.}(2012){Donley}, {Koekemoer}, {Brusa}, {Capak},
  {Cardamone}, {Civano}, {Ilbert}, {Impey}, {Kartaltepe}, {Miyaji}, {Salvato},
  {Sanders}, {Trump}, \& {Zamorani}}]{Donley12}
{Donley}, J.~L., {Koekemoer}, A.~M., {Brusa}, M., {et~al.} 2012, \apj, 748, 142

\bibitem[{{Draine} {et~al.}(2007){Draine}, {Dale}, {Bendo}, {Gordon}, {Smith},
  {Armus}, {Engelbracht}, {Helou}, {Kennicutt}, {Li}, {Roussel}, {Walter},
  {Calzetti}, {Moustakas}, {Murphy}, {Rieke}, {Bot}, {Hollenbach}, {Sheth}, \&
  {Teplitz}}]{Draine07}
{Draine}, B.~T., {Dale}, D.~A., {Bendo}, G., {et~al.} 2007, \apj, 663, 866

\bibitem[{{Dunlop} {et~al.}(2017){Dunlop}, {McLure}, {Biggs}, {Geach},
  {Micha{\l}owski}, {Ivison}, {Rujopakarn}, {van Kampen}, {Kirkpatrick},
  {Pope}, {Scott}, {Swinbank}, {Targett}, {Aretxaga}, {Austermann}, {Best},
  {Bruce}, {Chapin}, {Charlot}, {Cirasuolo}, {Coppin}, {Ellis}, {Finkelstein},
  {Hayward}, {Hughes}, {Ibar}, {Jagannathan}, {Khochfar}, {Koprowski},
  {Narayanan}, {Nyland}, {Papovich}, {Peacock}, {Rieke}, {Robertson},
  {Vernstrom}, {Werf}, {Wilson}, \& {Yun}}]{Dunlop17}
{Dunlop}, J.~S., {McLure}, R.~J., {Biggs}, A.~D., {et~al.} 2017, \mnras, 466,
  861

\bibitem[{{Dunne} {et~al.}(2000){Dunne}, {Eales}, {Edmunds}, {Ivison},
  {Alexander}, \& {Clements}}]{Dunne00}
{Dunne}, L., {Eales}, S., {Edmunds}, M., {et~al.} 2000, \mnras, 315, 115

\bibitem[{{Erb} {et~al.}(2006){Erb}, {Shapley}, {Pettini}, {Steidel}, {Reddy},
  \& {Adelberger}}]{Erb06a}
{Erb}, D.~K., {Shapley}, A.~E., {Pettini}, M., {et~al.} 2006, \apj, 644, 813

\bibitem[{{Feldmann}(2015)}]{Feldmann15}
{Feldmann}, R. 2015, \mnras, 449, 3274

\bibitem[{{Fujimoto} {et~al.}(2017){Fujimoto}, {Ouchi}, {Shibuya}, \&
  {Nagai}}]{Fujimoto17}
{Fujimoto}, S., {Ouchi}, M., {Shibuya}, T., {et~al.} 2017, ArXiv e-prints

\bibitem[{{Grogin} {et~al.}(2011){Grogin}, {Kocevski}, {Faber}, {Ferguson},
  {Koekemoer}, {Riess}, {Acquaviva}, {Alexander}, {Almaini}, {Ashby}, {Barden},
  {Bell}, {Bournaud}, {Brown}, {Caputi}, {Casertano}, {Cassata}, {Castellano},
  {Challis}, {Chary}, {Cheung}, {Cirasuolo}, {Conselice}, {Roshan Cooray},
  {Croton}, {Daddi}, {Dahlen}, {Dav{\'e}}, {de Mello}, {Dekel}, {Dickinson},
  {Dolch}, {Donley}, {Dunlop}, {Dutton}, {Elbaz}, {Fazio}, {Filippenko},
  {Finkelstein}, {Fontana}, {Gardner}, {Garnavich}, {Gawiser}, {Giavalisco},
  {Grazian}, {Guo}, {Hathi}, {H{\"a}ussler}, {Hopkins}, {Huang}, {Huang},
  {Jha}, {Kartaltepe}, {Kirshner}, {Koo}, {Lai}, {Lee}, {Li}, {Lotz}, {Lucas},
  {Madau}, {McCarthy}, {McGrath}, {McIntosh}, {McLure}, {Mobasher},
  {Moustakas}, {Mozena}, {Nandra}, {Newman}, {Niemi}, {Noeske}, {Papovich},
  {Pentericci}, {Pope}, {Primack}, {Rajan}, {Ravindranath}, {Reddy}, {Renzini},
  {Rix}, {Robaina}, {Rodney}, {Rosario}, {Rosati}, {Salimbeni}, {Scarlata},
  {Siana}, {Simard}, {Smidt}, {Somerville}, {Spinrad}, {Straughn}, {Strolger},
  {Telford}, {Teplitz}, {Trump}, {van der Wel}, {Villforth}, {Wechsler},
  {Weiner}, {Wiklind}, {Wild}, {Wilson}, {Wuyts}, {Yan}, \& {Yun}}]{Grogin11}
{Grogin}, N.~A., {Kocevski}, D.~D., {Faber}, S.~M., {et~al.} 2011, \apjs, 197,
  35

\bibitem[{{Heinis} {et~al.}(2014){Heinis}, {Buat}, {B{\'e}thermin}, {Bock},
  {Burgarella}, {Conley}, {Cooray}, {Farrah}, {Ilbert}, {Magdis}, {Marsden},
  {Oliver}, {Rigopoulou}, {Roehlly}, {Schulz}, {Symeonidis}, {Viero}, {Xu}, \&
  {Zemcov}}]{Heinis14}
{Heinis}, S., {Buat}, V., {B{\'e}thermin}, M., {et~al.} 2014, \mnras, 437, 1268

\bibitem[{{Hirashita}(2012)}]{Hirashita12}
{Hirashita}, H. 2012, \mnras, 422, 1263

\bibitem[{{Hirashita}(2015)}]{Hirashita15}
---. 2015, \mnras, 447, 2937

\bibitem[{{Iono} {et~al.}(2006){Iono}, {Peck}, {Pope}, {Borys}, {Scott},
  {Wilner}, {Gurwell}, {Ho}, {Yun}, {Matsushita}, {Petitpas}, {Dunlop},
  {Elvis}, {Blain}, \& {Le Floc'h}}]{Iono06}
{Iono}, D., {Peck}, A.~B., {Pope}, A., {et~al.} 2006, \apjl, 640, L1

\bibitem[{{Kauffmann} {et~al.}(2003){Kauffmann}, {Heckman}, {White}, {Charlot},
  {Tremonti}, {Brinchmann}, {Bruzual}, {Peng}, {Seibert}, {Bernardi},
  {Blanton}, {Brinkmann}, {Castander}, {Cs{\'a}bai}, {Fukugita}, {Ivezic},
  {Munn}, {Nichol}, {Padmanabhan}, {Thakar}, {Weinberg}, \&
  {York}}]{Kauffmann03}
{Kauffmann}, G., {Heckman}, T.~M., {White}, S.~D.~M., {et~al.} 2003, \mnras,
  341, 33

\bibitem[{{Kennicutt}(1998{\natexlab{a}})}]{Kennicutt98}
{Kennicutt}, Jr., R.~C. 1998{\natexlab{a}}, \araa, 36, 189

\bibitem[{{Kennicutt}(1998{\natexlab{b}})}]{Kennicutt98b}
---. 1998{\natexlab{b}}, \apj, 498, 541

\bibitem[{{Kennicutt} {et~al.}(2009){Kennicutt}, {Hao}, {Calzetti},
  {Moustakas}, {Dale}, {Bendo}, {Engelbracht}, {Johnson}, \&
  {Lee}}]{Kennicutt09}
{Kennicutt}, Jr., R.~C., {Hao}, C.-N., {Calzetti}, D., {et~al.} 2009, \apj,
  703, 1672

\bibitem[{{Kewley} \& {Ellison}(2008)}]{Kewley08}
{Kewley}, L.~J., \& {Ellison}, S.~L. 2008, \apj, 681, 1183

\bibitem[{{Kirkpatrick} {et~al.}(2012){Kirkpatrick}, {Pope}, {Alexander},
  {Charmandaris}, {Daddi}, {Dickinson}, {Elbaz}, {Gabor}, {Hwang}, {Ivison},
  {Mullaney}, {Pannella}, {Scott}, {Altieri}, {Aussel}, {Bournaud}, {Buat},
  {Coia}, {Dannerbauer}, {Dasyra}, {Kartaltepe}, {Leiton}, {Lin}, {Magdis},
  {Magnelli}, {Morrison}, {Popesso}, \& {Valtchanov}}]{Kirkpatrick12}
{Kirkpatrick}, A., {Pope}, A., {Alexander}, D.~M., {et~al.} 2012, \apj, 759,
  139

\bibitem[{{Kirkpatrick} {et~al.}(2017){Kirkpatrick}, {Pope}, {Sajina}, {Dale},
  {D{\'{\i}}az-Santos}, {Hayward}, {Shi}, {Somerville}, {Stierwalt}, {Armus},
  {Kartaltepe}, {Kocevski}, {McIntosh}, {Sanders}, \& {Yan}}]{Kirkpatrick17}
{Kirkpatrick}, A., {Pope}, A., {Sajina}, A., {et~al.} 2017, \apj, 843, 71

\bibitem[{{Kirkpatrick} {et~al.}(2015){Kirkpatrick}, {Pope}, {Sajina},
  {Roebuck}, {Yan}, {Armus}, {D{\'{\i}}az-Santos}, \&
  {Stierwalt}}]{Kirkpatrick15}
---. 2015, \apj, 814, 9

\bibitem[{{Koprowski} {et~al.}(2016){Koprowski}, {Coppin}, {Geach}, {Hine},
  {Bremer}, {Chapman}, {Davies}, {Hayashino}, {Knudsen}, {Kubo}, {Lehmer},
  {Matsuda}, {Smith}, {van der Werf}, {Violino}, \& {Yamada}}]{Koprowski16}
{Koprowski}, M.~P., {Coppin}, K.~E.~K., {Geach}, J.~E., {et~al.} 2016, \apjl,
  828, L21

\bibitem[{{Kriek} \& {Conroy}(2013)}]{Kriek13}
{Kriek}, M., \& {Conroy}, C. 2013, \apjl, 775, L16

\bibitem[{{Kriek} {et~al.}(2009){Kriek}, {van Dokkum}, {Franx}, {Illingworth},
  \& {Magee}}]{Kriek09b}
{Kriek}, M., {van Dokkum}, P.~G., {Franx}, M., {et~al.} 2009, \apjl, 705, L71

\bibitem[{{Leroy} {et~al.}(2011){Leroy}, {Bolatto}, {Gordon}, {Sandstrom},
  {Gratier}, {Rosolowsky}, {Engelbracht}, {Mizuno}, {Corbelli}, {Fukui}, \&
  {Kawamura}}]{Leroy11}
{Leroy}, A.~K., {Bolatto}, A., {Gordon}, K., {et~al.} 2011, \apj, 737, 12

\bibitem[{{Livermore} {et~al.}(2015){Livermore}, {Jones}, {Richard}, {Bower},
  {Swinbank}, {Yuan}, {Edge}, {Ellis}, {Kewley}, {Smail}, {Coppin}, \&
  {Ebeling}}]{Livermore15}
{Livermore}, R.~C., {Jones}, T.~A., {Richard}, J., {et~al.} 2015, \mnras, 450,
  1812

\bibitem[{{Madau} \& {Dickinson}(2014)}]{Madau14}
{Madau}, P., \& {Dickinson}, M. 2014, \araa, 52, 415

\bibitem[{{Magdis} {et~al.}(2012){Magdis}, {Daddi}, {B{\'e}thermin}, {Sargent},
  {Elbaz}, {Pannella}, {Dickinson}, {Dannerbauer}, {da Cunha}, {Walter},
  {Rigopoulou}, {Charmandaris}, {Hwang}, \& {Kartaltepe}}]{Magdis12}
{Magdis}, G.~E., {Daddi}, E., {B{\'e}thermin}, M., {et~al.} 2012, \apj, 760, 6

\bibitem[{{Magnelli} {et~al.}(2009){Magnelli}, {Elbaz}, {Chary}, {Dickinson},
  {Le Borgne}, {Frayer}, \& {Willmer}}]{Magnelli09}
{Magnelli}, B., {Elbaz}, D., {Chary}, R.~R., {et~al.} 2009, \aap, 496, 57

\bibitem[{{Martis} {et~al.}(2016){Martis}, {Marchesini}, {Brammer}, {Muzzin},
  {Labb{\'e}}, {Momcheva}, {Skelton}, {Stefanon}, {van Dokkum}, \&
  {Whitaker}}]{Martis16}
{Martis}, N.~S., {Marchesini}, D., {Brammer}, G.~B., {et~al.} 2016, \apjl, 827,
  L25

\bibitem[{{Meurer} {et~al.}(1999){Meurer}, {Heckman}, \& {Calzetti}}]{Meurer99}
{Meurer}, G.~R., {Heckman}, T.~M., \& {Calzetti}, D. 1999, \apj, 521, 64

\bibitem[{{Momcheva} {et~al.}(2016){Momcheva}, {Brammer}, {van Dokkum},
  {Skelton}, {Whitaker}, {Nelson}, {Fumagalli}, {Maseda}, {Leja}, {Franx},
  {Rix}, {Bezanson}, {Da Cunha}, {Dickey}, {F{\"o}rster Schreiber},
  {Illingworth}, {Kriek}, {Labb{\'e}}, {Ulf Lange}, {Lundgren}, {Magee},
  {Marchesini}, {Oesch}, {Pacifici}, {Patel}, {Price}, {Tal}, {Wake}, {van der
  Wel}, \& {Wuyts}}]{Momcheva15}
{Momcheva}, I.~G., {Brammer}, G.~B., {van Dokkum}, P.~G., {et~al.} 2016, \apjs,
  225, 27

\bibitem[{{Morokuma-Matsui} \& {Baba}(2015)}]{Morokuma15}
{Morokuma-Matsui}, K., \& {Baba}, J. 2015, \mnras, 454, 3792

\bibitem[{{Mu{\~n}oz-Mateos} {et~al.}(2009){Mu{\~n}oz-Mateos}, {Gil de Paz},
  {Boissier}, {Zamorano}, {Dale}, {P{\'e}rez-Gonz{\'a}lez}, {Gallego},
  {Madore}, {Bendo}, {Thornley}, {Draine}, {Boselli}, {Buat}, {Calzetti},
  {Moustakas}, \& {Kennicutt}}]{MunozMateos09}
{Mu{\~n}oz-Mateos}, J.~C., {Gil de Paz}, A., {Boissier}, S., {et~al.} 2009,
  \apj, 701, 1965

\bibitem[{{Murphy} {et~al.}(2011){Murphy}, {Condon}, {Schinnerer}, {Kennicutt},
  {Calzetti}, {Armus}, {Helou}, {Turner}, {Aniano}, {Beir{\~a}o}, {Bolatto},
  {Brandl}, {Croxall}, {Dale}, {Donovan Meyer}, {Draine}, {Engelbracht},
  {Hunt}, {Hao}, {Koda}, {Roussel}, {Skibba}, \& {Smith}}]{Murphy11}
{Murphy}, E.~J., {Condon}, J.~J., {Schinnerer}, E., {et~al.} 2011, \apj, 737,
  67

\bibitem[{{Muzzin} {et~al.}(2013){Muzzin}, {Marchesini}, {Stefanon}, {Franx},
  {Milvang-Jensen}, {Dunlop}, {Fynbo}, {Brammer}, {Labb{\'e}}, \& {van
  Dokkum}}]{Muzzin13}
{Muzzin}, A., {Marchesini}, D., {Stefanon}, M., {et~al.} 2013, \apjs, 206, 8

\bibitem[{{Narayanan} {et~al.}(2015){Narayanan}, {Turk}, {Feldmann},
  {Robitaille}, {Hopkins}, {Thompson}, {Hayward}, {Ball},
  {Faucher-Gigu{\`e}re}, \& {Kere{\v s}}}]{Narayanan15}
{Narayanan}, D., {Turk}, M., {Feldmann}, R., {et~al.} 2015, \nat, 525, 496

\bibitem[{{Pannella} {et~al.}(2009){Pannella}, {Carilli}, {Daddi}, {McCracken},
  {Owen}, {Renzini}, {Strazzullo}, {Civano}, {Koekemoer}, {Schinnerer},
  {Scoville}, {Smol{\v c}i{\'c}}, {Taniguchi}, {Aussel}, {Kneib}, {Ilbert},
  {Mellier}, {Salvato}, {Thompson}, \& {Willott}}]{Pannella09}
{Pannella}, M., {Carilli}, C.~L., {Daddi}, E., {et~al.} 2009, \apjl, 698, L116

\bibitem[{{Pope} {et~al.}(2017){Pope}, {Monta{\~n}a}, {Battisti}, {Limousin},
  {Marchesini}, {Wilson}, {Alberts}, {Aretxaga}, {Avila-Reese}, {Ram{\'o}n
  Bermejo-Climent}, {Brammer}, {Bravo-Alfaro}, {Calzetti}, {Chary}, {Cybulski},
  {Giavalisco}, {Hughes}, {Kado-Fong}, {Keller}, {Kirkpatrick}, {Labbe},
  {Lange-Vagle}, {Lowenthal}, {Murphy}, {Oesch}, {Rosa Gonzalez},
  {S{\'a}nchez-Arg{\"u}elles}, {Shipley}, {Stefanon}, {Vega}, {Whitaker},
  {Williams}, {Yun}, {Zavala}, \& {Zeballos}}]{Pope17}
{Pope}, A., {Monta{\~n}a}, A., {Battisti}, A., {et~al.} 2017, \apj, 838, 137

\bibitem[{{Popping} {et~al.}(2015){Popping}, {Caputi}, {Trager}, {Somerville},
  {Dekel}, {Kassin}, {Kocevski}, {Koekemoer}, {Faber}, {Ferguson}, {Galametz},
  {Grogin}, {Guo}, {Lu}, {Wel}, \& {Weiner}}]{Popping15}
{Popping}, G., {Caputi}, K.~I., {Trager}, S.~C., {et~al.} 2015, \mnras, 454,
  2258

\bibitem[{{Popping} {et~al.}(2017){Popping}, {Puglisi}, \&
  {Norman}}]{Popping17b}
{Popping}, G., {Puglisi}, A., \& {Norman}, C.~A. 2017, ArXiv e-prints

\bibitem[{{Popping} {et~al.}(2016){Popping}, {Somerville}, \&
  {Galametz}}]{Popping17}
{Popping}, G., {Somerville}, R.~S., \& {Galametz}, M. 2016, ArXiv e-prints

\bibitem[{{Reddy} {et~al.}(2015){Reddy}, {Kriek}, {Shapley}, {Freeman},
  {Siana}, {Coil}, {Mobasher}, {Price}, {Sanders}, \& {Shivaei}}]{Reddy15}
{Reddy}, N.~A., {Kriek}, M., {Shapley}, A.~E., {et~al.} 2015, \apj, 806, 259

\bibitem[{{R{\'e}my-Ruyer} {et~al.}(2014){R{\'e}my-Ruyer}, {Madden},
  {Galliano}, {Galametz}, {Takeuchi}, {Asano}, {Zhukovska}, {Lebouteiller},
  {Cormier}, {Jones}, {Bocchio}, {Baes}, {Bendo}, {Boquien}, {Boselli},
  {DeLooze}, {Doublier-Pritchard}, {Hughes}, {Karczewski}, \&
  {Spinoglio}}]{Remy14}
{R{\'e}my-Ruyer}, A., {Madden}, S.~C., {Galliano}, F., {et~al.} 2014, \aap,
  563, A31

\bibitem[{{Rodighiero} {et~al.}(2011){Rodighiero}, {Daddi}, {Baronchelli},
  {Cimatti}, {Renzini}, {Aussel}, {Popesso}, {Lutz}, {Andreani}, {Berta},
  {Cava}, {Elbaz}, {Feltre}, {Fontana}, {F{\"o}rster Schreiber},
  {Franceschini}, {Genzel}, {Grazian}, {Gruppioni}, {Ilbert}, {Le Floch},
  {Magdis}, {Magliocchetti}, {Magnelli}, {Maiolino}, {McCracken}, {Nordon},
  {Poglitsch}, {Santini}, {Pozzi}, {Riguccini}, {Tacconi}, {Wuyts}, \&
  {Zamorani}}]{Rodighiero11}
{Rodighiero}, G., {Daddi}, E., {Baronchelli}, I., {et~al.} 2011, \apjl, 739,
  L40

\bibitem[{{Saintonge} {et~al.}(2016){Saintonge}, {Catinella}, {Cortese},
  {Genzel}, {Giovanelli}, {Haynes}, {Janowiecki}, {Kramer}, {Lutz},
  {Schiminovich}, {Tacconi}, {Wuyts}, \& {Accurso}}]{Saintonge16}
{Saintonge}, A., {Catinella}, B., {Cortese}, L., {et~al.} 2016, \mnras, 462,
  1749

\bibitem[{{Saintonge} {et~al.}(2011){Saintonge}, {Kauffmann}, {Kramer},
  {Tacconi}, {Buchbender}, {Catinella}, {Fabello}, {Graci{\'a}-Carpio}, {Wang},
  {Cortese}, {Fu}, {Genzel}, {Giovanelli}, {Guo}, {Haynes}, {Heckman},
  {Krumholz}, {Lemonias}, {Li}, {Moran}, {Rodriguez-Fernandez}, {Schiminovich},
  {Schuster}, \& {Sievers}}]{Saintonge11}
{Saintonge}, A., {Kauffmann}, G., {Kramer}, C., {et~al.} 2011, \mnras, 415, 32

\bibitem[{{Salmon} {et~al.}(2016){Salmon}, {Papovich}, {Long}, {Willner},
  {Finkelstein}, {Ferguson}, {Dickinson}, {Duncan}, {Faber}, {Hathi},
  {Koekemoer}, {Kurczynski}, {Newman}, {Pacifici}, {P{\'e}rez-Gonz{\'a}lez}, \&
  {Pforr}}]{Salmon16}
{Salmon}, B., {Papovich}, C., {Long}, J., {et~al.} 2016, \apj, 827, 20

\bibitem[{{Sanders} {et~al.}(2007){Sanders}, {Salvato}, {Aussel}, {Ilbert},
  {Scoville}, {Surace}, {Frayer}, {Sheth}, {Helou}, {Brooke}, {Bhattacharya},
  {Yan}, {Kartaltepe}, {Barnes}, {Blain}, {Calzetti}, {Capak}, {Carilli},
  {Carollo}, {Comastri}, {Daddi}, {Ellis}, {Elvis}, {Fall}, {Franceschini},
  {Giavalisco}, {Hasinger}, {Impey}, {Koekemoer}, {Le F{\`e}vre}, {Lilly},
  {Liu}, {McCracken}, {Mobasher}, {Renzini}, {Rich}, {Schinnerer}, {Shopbell},
  {Taniguchi}, {Thompson}, {Urry}, \& {Williams}}]{Sanders07}
{Sanders}, D.~B., {Salvato}, M., {Aussel}, H., {et~al.} 2007, \apjs, 172, 86

\bibitem[{{Sanders} {et~al.}(2015){Sanders}, {Shapley}, {Kriek}, {Reddy},
  {Freeman}, {Coil}, {Siana}, {Mobasher}, {Shivaei}, {Price}, \& {de
  Groot}}]{Sanders15}
{Sanders}, R.~L., {Shapley}, A.~E., {Kriek}, M., {et~al.} 2015, \apj, 799, 138

\bibitem[{{Santini} {et~al.}(2014){Santini}, {Maiolino}, {Magnelli}, {Lutz},
  {Lamastra}, {Li Causi}, {Eales}, {Andreani}, {Berta}, {Buat}, {Cooray},
  {Cresci}, {Daddi}, {Farrah}, {Fontana}, {Franceschini}, {Genzel}, {Granato},
  {Grazian}, {Le Floc'h}, {Magdis}, {Magliocchetti}, {Mannucci}, {Menci},
  {Nordon}, {Oliver}, {Popesso}, {Pozzi}, {Riguccini}, {Rodighiero}, {Rosario},
  {Salvato}, {Scott}, {Silva}, {Tacconi}, {Viero}, {Wang}, {Wuyts}, \&
  {Xu}}]{Santini14}
{Santini}, P., {Maiolino}, R., {Magnelli}, B., {et~al.} 2014, \aap, 562, A30

\bibitem[{{Savaglio} {et~al.}(2005){Savaglio}, {Glazebrook}, {Le Borgne},
  {Juneau}, {Abraham}, {Chen}, {Crampton}, {McCarthy}, {Carlberg}, {Marzke},
  {Roth}, {J{\o}rgensen}, \& {Murowinski}}]{Savaglio05}
{Savaglio}, S., {Glazebrook}, K., {Le Borgne}, D., {et~al.} 2005, \apj, 635,
  260

\bibitem[{{Schmidt}(1959)}]{Schmidt59}
{Schmidt}, M. 1959, \apj, 129, 243

\bibitem[{{Schreiber} {et~al.}(2015){Schreiber}, {Pannella}, {Elbaz},
  {B{\'e}thermin}, {Inami}, {Dickinson}, {Magnelli}, {Wang}, {Aussel}, {Daddi},
  {Juneau}, {Shu}, {Sargent}, {Buat}, {Faber}, {Ferguson}, {Giavalisco},
  {Koekemoer}, {Magdis}, {Morrison}, {Papovich}, {Santini}, \&
  {Scott}}]{Schreiber15}
{Schreiber}, C., {Pannella}, M., {Elbaz}, D., {et~al.} 2015, \aap, 575, A74

\bibitem[{{Scoville} {et~al.}(2016){Scoville}, {Sheth}, {Aussel}, {Vanden
  Bout}, {Capak}, {Bongiorno}, {Casey}, {Murchikova}, {Koda},
  {{\'A}lvarez-M{\'a}rquez}, {Lee}, {Laigle}, {McCracken}, {Ilbert}, {Pope},
  {Sanders}, {Chu}, {Toft}, {Ivison}, \& {Manohar}}]{Scoville16}
{Scoville}, N., {Sheth}, K., {Aussel}, H., {et~al.} 2016, \apj, 820, 83

\bibitem[{{Seibert} {et~al.}(2005){Seibert}, {Martin}, {Heckman}, {Buat},
  {Hoopes}, {Barlow}, {Bianchi}, {Byun}, {Donas}, {Forster}, {Friedman},
  {Jelinsky}, {Lee}, {Madore}, {Malina}, {Milliard}, {Morrissey}, {Neff},
  {Rich}, {Schiminovich}, {Siegmund}, {Small}, {Szalay}, {Welsh}, \&
  {Wyder}}]{Seibert05}
{Seibert}, M., {Martin}, D.~C., {Heckman}, T.~M., {et~al.} 2005, \apjl, 619,
  L55

\bibitem[{{Skelton} {et~al.}(2014){Skelton}, {Whitaker}, {Momcheva}, {Brammer},
  {van Dokkum}, {Labb{\'e}}, {Franx}, {van der Wel}, {Bezanson}, {Da Cunha},
  {Fumagalli}, {F{\"o}rster Schreiber}, {Kriek}, {Leja}, {Lundgren}, {Magee},
  {Marchesini}, {Maseda}, {Nelson}, {Oesch}, {Pacifici}, {Patel}, {Price},
  {Rix}, {Tal}, {Wake}, \& {Wuyts}}]{Skelton14}
{Skelton}, R.~E., {Whitaker}, K.~E., {Momcheva}, I.~G., {et~al.} 2014, \apjs,
  214, 24

\bibitem[{{Sobral} {et~al.}(2012){Sobral}, {Best}, {Matsuda}, {Smail}, {Geach},
  \& {Cirasuolo}}]{Sobral12}
{Sobral}, D., {Best}, P.~N., {Matsuda}, Y., {et~al.} 2012, \mnras, 420, 1926

\bibitem[{{Speagle} {et~al.}(2014){Speagle}, {Steinhardt}, {Capak}, \&
  {Silverman}}]{Speagle14}
{Speagle}, J.~S., {Steinhardt}, C.~L., {Capak}, P.~L., {et~al.} 2014, ArXiv
  e-prints

\bibitem[{{Steidel} {et~al.}(2014){Steidel}, {Rudie}, {Strom}, {Pettini},
  {Reddy}, {Shapley}, {Trainor}, {Erb}, {Turner}, {Konidaris}, {Kulas}, {Mace},
  {Matthews}, \& {McLean}}]{Steidel14}
{Steidel}, C.~C., {Rudie}, G.~C., {Strom}, A.~L., {et~al.} 2014, ArXiv e-prints

\bibitem[{{Tacconi} {et~al.}(2017){Tacconi}, {Genzel}, {Saintonge}, {Combes},
  {Garc{\'{\i}}a-Burillo}, {Neri}, {Bolatto}, {Contini}, {F{\"o}rster
  Schreiber}, {Lilly}, {Lutz}, {Wuyts}, {Accurso}, {Boissier}, {Boone},
  {Bouch{\'e}}, {Bournaud}, {Burkert}, {Carollo}, {Cooper}, {Cox}, {Feruglio},
  {Freundlich}, {Herrera-Camus}, {Juneau}, {Lippa}, {Naab}, {Renzini},
  {Salome}, {Sternberg}, {Tadaki}, {{\"U}bler}, {Walter}, {Weiner}, \&
  {Weiss}}]{Tacconi17}
{Tacconi}, L.~J., {Genzel}, R., {Saintonge}, A., {et~al.} 2017, ArXiv e-prints

\bibitem[{{Tacconi} {et~al.}(2013){Tacconi}, {Neri}, {Genzel}, {Combes},
  {Bolatto}, {Cooper}, {Wuyts}, {Bournaud}, {Burkert}, {Comerford}, {Cox},
  {Davis}, {F{\"o}rster Schreiber}, {Garc{\'{\i}}a-Burillo}, {Gracia-Carpio},
  {Lutz}, {Naab}, {Newman}, {Omont}, {Saintonge}, {Shapiro Griffin}, {Shapley},
  {Sternberg}, \& {Weiner}}]{Tacconi13}
{Tacconi}, L.~J., {Neri}, R., {Genzel}, R., {et~al.} 2013, \apj, 768, 74

\bibitem[{{Tal} {et~al.}(2014){Tal}, {Dekel}, {Oesch}, {Muzzin}, {Brammer},
  {van Dokkum}, {Franx}, {Illingworth}, {Leja}, {Magee}, {Marchesini},
  {Momcheva}, {Nelson}, {Patel}, {Quadri}, {Rix}, {Skelton}, {Wake}, \&
  {Whitaker}}]{Tal14}
{Tal}, T., {Dekel}, A., {Oesch}, P., {et~al.} 2014, ArXiv e-prints

\bibitem[{{Tomczak} {et~al.}(2016){Tomczak}, {Quadri}, {Tran}, {Labb{\'e}},
  {Straatman}, {Papovich}, {Glazebrook}, {Allen}, {Brammer}, {Cowley},
  {Dickinson}, {Elbaz}, {Inami}, {Kacprzak}, {Morrison}, {Nanayakkara},
  {Persson}, {Rees}, {Salmon}, {Schreiber}, {Spitler}, \&
  {Whitaker}}]{Tomczak16}
{Tomczak}, A.~R., {Quadri}, R.~F., {Tran}, K.-V.~H., {et~al.} 2016, \apj, 817,
  118

\bibitem[{{Utomo} {et~al.}(2014){Utomo}, {Kriek}, {Labb{\'e}}, {Conroy}, \&
  {Fumagalli}}]{Utomo14}
{Utomo}, D., {Kriek}, M., {Labb{\'e}}, I., {et~al.} 2014, \apjl, 783, L30

\bibitem[{{van der Wel} {et~al.}(2014){van der Wel}, {Franx}, {van Dokkum},
  {Skelton}, {Momcheva}, {Whitaker}, {Brammer}, {Bell}, {Rix}, {Wuyts},
  {Ferguson}, {Holden}, {Barro}, {Koekemoer}, {Chang}, {McGrath},
  {H{\"a}ussler}, {Dekel}, {Behroozi}, {Fumagalli}, {Leja}, {Lundgren},
  {Maseda}, {Nelson}, {Wake}, {Patel}, {Labb{\'e}}, {Faber}, {Grogin}, \&
  {Kocevski}}]{vanderWel14}
{van der Wel}, A., {Franx}, M., {van Dokkum}, P.~G., {et~al.} 2014, \apj, 788,
  28

\bibitem[{{van Dokkum} {et~al.}(2013){van Dokkum}, {Leja}, {Nelson}, {Patel},
  {Skelton}, {Momcheva}, {Brammer}, {Whitaker}, {Lundgren}, {Fumagalli},
  {Conroy}, {F{\"o}rster Schreiber}, {Franx}, {Kriek}, {Labb{\'e}},
  {Marchesini}, {Rix}, {van der Wel}, \& {Wuyts}}]{vanDokkum13}
{van Dokkum}, P.~G., {Leja}, J., {Nelson}, E.~J., {et~al.} 2013, \apjl, 771,
  L35

\bibitem[{{Viero} {et~al.}(2013){Viero}, {Moncelsi}, {Quadri}, {Arumugam},
  {Assef}, {B{\'e}thermin}, {Bock}, {Bridge}, {Casey}, {Conley}, {Cooray},
  {Farrah}, {Glenn}, {Heinis}, {Ibar}, {Ikarashi}, {Ivison}, {Kohno},
  {Marsden}, {Oliver}, {Roseboom}, {Schulz}, {Scott}, {Serra}, {Vaccari},
  {Vieira}, {Wang}, {Wardlow}, {Wilson}, {Yun}, \& {Zemcov}}]{Viero13}
{Viero}, M.~P., {Moncelsi}, L., {Quadri}, R.~F., {et~al.} 2013, \apj, 779, 32

\bibitem[{{Whitaker} {et~al.}(2017){Whitaker}, {Bezanson}, {van Dokkum},
  {Franx}, {van der Wel}, {Brammer}, {F{\"o}rster-Schreiber}, {Giavalisco},
  {Labb{\'e}}, {Momcheva}, {Nelson}, \& {Skelton}}]{Whitaker17}
{Whitaker}, K.~E., {Bezanson}, R., {van Dokkum}, P.~G., {et~al.} 2017, \apj,
  838, 19

\bibitem[{{Whitaker} {et~al.}(2014){Whitaker}, {Franx}, {Leja}, {van Dokkum},
  {Henry}, {Skelton}, {Fumagalli}, {Momcheva}, {Brammer}, {Labb{\'e}},
  {Nelson}, \& {Rigby}}]{Whitaker14b}
{Whitaker}, K.~E., {Franx}, M., {Leja}, J., {et~al.} 2014, \apj, 795, 104

\bibitem[{{Whitaker} {et~al.}(2012{\natexlab{a}}){Whitaker}, {Kriek}, {van
  Dokkum}, {Bezanson}, {Brammer}, {Franx}, \& {Labb{\'e}}}]{Whitaker12a}
{Whitaker}, K.~E., {Kriek}, M., {van Dokkum}, P.~G., {et~al.}
  2012{\natexlab{a}}, \apj, 745, 179

\bibitem[{{Whitaker} {et~al.}(2012{\natexlab{b}}){Whitaker}, {van Dokkum},
  {Brammer}, \& {Franx}}]{Whitaker12b}
{Whitaker}, K.~E., {van Dokkum}, P.~G., {Brammer}, G., {et~al.}
  2012{\natexlab{b}}, \apjl, 754, L29

\bibitem[{{Wild} {et~al.}(2011){Wild}, {Charlot}, {Brinchmann}, {Heckman},
  {Vince}, {Pacifici}, \& {Chevallard}}]{Wild11}
{Wild}, V., {Charlot}, S., {Brinchmann}, J., {et~al.} 2011, \mnras, 417, 1760

\bibitem[{{Wisnioski} {et~al.}(2012){Wisnioski}, {Glazebrook}, {Blake},
  {Poole}, {Green}, {Wyder}, \& {Martin}}]{Wisnioski12}
{Wisnioski}, E., {Glazebrook}, K., {Blake}, C., {et~al.} 2012, \mnras, 422,
  3339

\bibitem[{{Wuyts} {et~al.}(2011){Wuyts}, {F{\"o}rster Schreiber}, {Lutz},
  {Nordon}, {Berta}, {Altieri}, {Andreani}, {Aussel}, {Bongiovanni}, {Cepa},
  {Cimatti}, {Daddi}, {Elbaz}, {Genzel}, {Koekemoer}, {Magnelli}, {Maiolino},
  {McGrath}, {P{\'e}rez Garc{\'{\i}}a}, {Poglitsch}, {Popesso}, {Pozzi},
  {Sanchez-Portal}, {Sturm}, {Tacconi}, \& {Valtchanov}}]{Wuyts11a}
{Wuyts}, S., {F{\"o}rster Schreiber}, N.~M., {Lutz}, D., {et~al.} 2011, \apj,
  738, 106

\bibitem[{{Zahid} {et~al.}(2012){Zahid}, {Dima}, {Kewley}, {Erb}, \&
  {Dav{\'e}}}]{Zahid12}
{Zahid}, H.~J., {Dima}, G.~I., {Kewley}, L.~J., {et~al.} 2012, \apj, 757, 54

\bibitem[{{Zahid} {et~al.}(2011){Zahid}, {Kewley}, \& {Bresolin}}]{Zahid11}
{Zahid}, H.~J., {Kewley}, L.~J., \& {Bresolin}, F. 2011, \apj, 730, 137

\bibitem[{{Zeimann} {et~al.}(2015){Zeimann}, {Ciardullo}, {Gronwall}, {Bridge},
  {Brooks}, {Fox}, {Gawiser}, {Gebhardt}, {Hagen}, {Schneider}, \&
  {Trump}}]{Zeimann15}
{Zeimann}, G.~R., {Ciardullo}, R., {Gronwall}, C., {et~al.} 2015, \apj, 814,
  162

\end{thebibliography}
\end{document}